\documentclass[modern]{aastex62}
\usepackage{ae,aecompl}
\usepackage{epsfig}
\usepackage{amsmath}
\usepackage{graphicx}
\usepackage{url}
\usepackage{amssymb}
\usepackage{times}
\usepackage{float}
\usepackage{natbib}
\usepackage{multirow}
\usepackage{booktabs}

\def\beq{\begin{equation}}
\def\eeq{\end{equation}}
\def\beqar{\begin{eqnarray}}
\def\eeqar{\end{eqnarray}}

\def\avg#1{\langle #1 \rangle}

\def\msol{M_\odot}

\def\isotope#1#2{\mbox{${}^{#2}{\rm #1}$}}
\def\fe5#1{\isotope{Fe}{5#1}}
\def\co5#1{\isotope{Co}{5#1}}
\def\ni5#1{\isotope{Ni}{5#1}}

\def\apj{ApJ}
\def\aap{A$\&$A}
\def\apjl{ApJL}
\def\apjs{ApJS}
\def\nat{Nature}

\def\aj{The Astronomical Journal}

\def\mnras{MNRAS}
\def\prd{PhRvD}

\def\fun#1#2{\lower3.6pt\vbox{\baselineskip0pt\lineskip.9pt
 \ialign{$\mathsurround=0pt#1\hfil##\hfil$\crcr#2\crcr\sim\crcr}}}

\submitjournal{ApJ}

\shorttitle{Spallation of \textit{r}-Process Nuclei}
\shortauthors{Wang et al.}

\begin{document}

\graphicspath{{./}{figures/}}

\title{Sandblasting the \textit{r}-Process:  Spallation of Ejecta from Neutron Star Mergers}

\correspondingauthor{Xilu Wang}
\email{xwang50@nd.edu, xlwang811@gmail.com}

\author[0000-0002-0786-7307]{Xilu Wang}
\altaffiliation{N3AS Postdoctoral Fellow}
\affil{Department of Physics, University of California, Berkeley, CA 94720, USA}
\affil{Department of Physics, University of Notre Dame, Notre Dame, IN 46556, USA}
\collaboration{(N3AS collaboration)}

\author{Brian D. Fields}
\affiliation{Department of Astronomy, University of Illinois at Urbana-Champaign, Urbana, IL 61801, USA}
\affiliation{Department of Physics, University of Illinois at Urbana-Champaign, Urbana, IL 61801, USA}

\author{Matthew Mumpower}
\affiliation{Theoretical Division, Los Alamos National Laboratory, Los Alamos, NM, 87545, USA}
\affiliation{Center for Theoretical Astrophysics, Los Alamos National Laboratory, Los Alamos, NM, 87545, USA}

\author{Trevor Sprouse}
\affil{Department of Physics, University of Notre Dame, Notre Dame, IN 46556, USA}
\affil{Theoretical Division, Los Alamos National Laboratory, Los Alamos, NM, 87545, USA}

\author{Rebecca Surman}
\affil{Department of Physics, University of Notre Dame, Notre Dame, IN 46556, USA}

\author{Nicole Vassh}
\affil{Department of Physics, University of Notre Dame, Notre Dame, IN 46556, USA}

\begin{abstract}

Neutron star mergers (NSMs) are rapid neutron capture (\textit{r}-process) nucleosynthesis sites that expel matter at high velocities, from $0.1c$ to as high as $0.6c$.
Nuclei ejected at these speeds are sufficiently energetic to initiate spallation nuclear reactions with interstellar medium (ISM) particles. We adopt a thick-target model for the propagation of high-speed heavy nuclei in the ISM, similar to the transport of cosmic rays.
We find that spallation may create observable perturbations to NSM isotopic abundances, particularly around the low-mass edges of the \textit{r}-process peaks where neighboring nuclei have very different abundances.
The extent to which spallation modifies the final NSM isotopic yields depends on:  (1) the ejected abundances, which are determined by the NSM astrophysical conditions and the properties of nuclei far from stability, (2) the ejecta velocity distribution and propagation in interstellar matter, and (3) the spallation cross sections.
Observed solar and stellar \textit{r}-process yields could thus constrain the velocity distribution of ejected
neutron star matter, assuming NSMs are the dominant \textit{r}-process source. We suggest avenues for future work, including measurement of relevant
cross sections.
\end{abstract}

\keywords{cosmic rays, \textit{r}-process, nucleosynthesis, nuclear reaction cross sections, nuclear abundances, compact binary stars}

\section{Introduction} 
\label{sec:intro}

When particles travel with high speeds ($v\gtrsim 0.1c$; kinetic energy $E_{k}\gtrsim 5$ MeV) in the interstellar medium (ISM), they are sufficiently energetic to interact with the ISM particles by nuclear reactions. 
These reactions generally lead to {\em spallation}: fragmentation processes in which a heavy nucleus emits one or more nucleons, thus reducing its atomic weight. 

Nuclear spallation is well studied in the context of cosmic rays. Cosmic rays are highly enriched in Li, Be, and B relative to the ISM or solar system, due to fragmentation of cosmic ray C, N, and O nuclei during propagation in the interstellar medium \citep[e.g.,][]{George2009}. The effect of spallation on the cosmic-ray abundance pattern is thus to ``fill in the valley'' at Li, Be, and B, at the expense of a small reduction in the neighboring CNO peak. Indeed, this effect is not only important for cosmic-ray abundances, but is an important and sometimes dominant nucleosynthesis source for Li, Be, and B generally \citep{Reeves1970,CR1,Walker1985,Duncan1992,Fields1994, Higdon1998, Lemoine1998, Fields2000, Ramaty2000, Suzuki2001}. 

Spallation may also affect the nucleosynthetic outcomes of individual energetic astrophysical events. For example, spallation reactions can occur in the fast ejecta from supernovae or hypernovae as it interacts with the circumstellar medium \citep[e.g.,][]{Fields2002, Nakamura2004}. Neutron star mergers (NSMs) are another potential site of interest for spallation studies. Spallation reactions influence the composition of the ultra-heavy cosmic rays \citep[e.g.,][]{Binns2019} that owe their origins to NSMs \citep{Komiya2017}. They could also alter the abundances of nuclei synthesized in the NSM event, as fast outflowing material from the merger first encounters the ISM. The latter has not yet been considered and is our focus here.

The bulk of heavy nuclei ejected from NSMs are expected to be synthesized through the rapid neutron-capture process (\textit{r}-process), which is one dominant nucleosynthesis avenue for heavy elements, especially for those heavier than the iron group \citep{Burbidge1957, Cameron1957}. In the \textit{r} process, rapid neutron capture pushes material far from stability and shapes the characteristic abundance pattern with three distinct peaks (at mass numbers $A\sim80$,  $A\sim130$, and $A\sim196$), associated with closed shell structures (at neutron numbers $N=50$, $N=82$, and $N=126$). These peaks are clearly seen in the abundance pattern of our solar system \citep[e.g.,][]{Lodders2003, Sneden2008}, where approximately half of the heavy elements have an \textit{r}-process origin. While the astrophysical site(s) responsible for the galactic tally of \textit{r}-process elements are still uncertain \citep[see reviews][and references therein]{Cowan1991, solar2007, Cowan2019, Kajino2019}, NSMs are the first verified site of \textit{r}-process nucleosynthesis \citep{Abbott, NSM}. The kilonova signal from the multi-messenger event GW170817 indicated lanthanide production from an NSM \citep{Cowperthwaite2017, Kasen2017}.

In addition, kilonova models of the GW170817 observation suggested that NSMs eject \textit{r}-process material with high speed that ranges from $0.1c$ to $0.3c$ on average \citep[e.g.,][]{Kasen2017, Rosswog2018, Wollaeger2018, Watson2019}, as expected from previous theoretical work \citep[e.g.,][]{Li1998, Tanaka2013}. {Observations of gamma ray bursts, which may potentially be associated with kilonovae, also indicate similar velocity ranges $\sim 0.1c - 0.7c$ \citep[e.g.,][]{GRB1,GRB2,GRB3,Jin2019}}. Thus, as this ejecta interacts with interstellar medium, spallation reactions can occur and may influence the overall nucleosynthesis yields from the event.  These interactions are the first steps in the NSM ejecta deceleration and, thus, are guaranteed to occur for any NSM matter that is eventually stopped and incorporated into observable systems such as stars or the presolar matter.

The \textit{r}-process ejecta speed and composition are the critical factors for the spallation process, and these depend on the ejecta origin. Simulations find that NSMs permit at least two distinct environments for heavy element production: 
(1) dynamical ejecta, which is expected to be very neutron-rich (electron fraction $Y_e\sim0.03-0.2$) and have high speeds in the subrelativistic regime $\sim0.1-0.3c$, with a velocity tail that can extend to  $\sim0.5-0.6c$ \citep{Bauswein2013, Hotokezaka2013, Rosswog2013, Endrizzi2016, Lehner2016, Sekiguchi2016, Rosswog2017}; 
(2) a viscous and/or neutrino-driven wind, which is believed to have a range of neutron-richness ($Y_e\sim0.2-0.5$), but lower velocity ($v\sim0.1c$) and mass \citep{Chen2007, Surman2008, Dessart2009, Perego2014, Wanajo2014, Just2015, Martin2015, Siegel2018}. Note that the conditions present in neutrino-driven winds are particularly uncertain due to the difficulties in treating neutrinos properly in such environments \citep[e.g.,][]{Caballero2012, Foucart2015, Malkus2016, Kyutoku2018}.

In this paper, we investigate the effect of spallation on the shapes of the \textit{r}-process abundance peaks produced in fast ejecta from an NSM event, and test whether spallation could alleviate the mismatch between simulation results and solar data.  In doing so, we explore the impact of the
many uncertainties in \textit{r}-process nucleosynthesis.  These include nuclear inputs for unstable nuclei and astrophysical conditions of the merger event \citep[e.g.,][]{Matt2016, Kajino2017}.  These uncertainties write themselves into the \textit{r}-process abundance patterns as
large variations around the second peak ($A\sim130$) and third peak ($A\sim195$). For example, the \textit{r}-process peaks can be narrower and shifted relative to solar data for some astrophysical conditions and choices of input nuclear physics \citep[e.g.,][]{Goriely2011, Korobkin2012, Just2015, Rosswog2017}.

The structure of this paper is as follows. Section~\ref{sec: method} describes the methods of our spallation calculation, including the thick-target model for heavy element transport, and the spallation cross sections of \textit{r}-process nuclei.
Section~\ref{sec: results} presents the spallation results for \textit{r}-process abundances calculated with different astrophysical and nuclear physics inputs. We identify the spallation cross sections with the greatest potential influence on the third peak ($A\sim195$) region in Section~\ref{subsec: sensitivity} and conclude in Section~\ref{sec: discussion}.

\section{Method}
\label{sec: method}

To calculate the potential influence of spallation on \textit{r}-process abundance patterns, we first generate initial abundance patterns of \textit{r}-process nuclei ejected from an NSM using the nucleosynthesis network code PRISM (see Section~\ref{sec: results}). We then adopt a thick-target model for propagation of the \textit{r}-process nuclei through the ISM, obtaining new abundances of the \textit{r}-process nuclei after spallation. 
{In this paper, we use the phrasing `initial \textit{r}-process abundance pattern' to indicate the isotopic pattern of abundances in the material ejected by the NSM event, prior to the onset of spallation reactions between it and the interstellar medium.}

\subsection{Model Assumptions}
\label{subsec: assumptions}
 
We describe the propagation of the \textit{r}-process nuclei in a one-zone, thick-target ``closed-box" model. The initial conditions for our model consist of \textit{r}-process ejecta from an NSM event.  
These heavy nuclei propagate into an ISM composed of hydrogen and $^{4}{\rm He}$ and experience ionization loss and spallation reactions. The spallation reactions are responsible for the change in the abundance pattern of heavy nuclei as the NSM ejecta is thermalized and eventually incorporated into future generations of stars.

The basic assumptions of our model are as follows:\\
\begin{enumerate}
\item
The NSM outflows have initial abundances that depend on the adopted \textit{r}-process nucleosynthesis model and are ejected with a uniform initial speed $v$.

\item
The NSM ejecta and ISM are both spatially homogeneous.

\item
All of the \textit{r}-process nuclei will interact with the ISM, i.e., the escape rate is zero. Advection and diffusion loss are also ignored here. 

\item
Among the energy loss mechanisms, ionic/Coulomb losses dominate. For \textit{r}-process nuclei with similar mass number $A$ and charge number $Z$, the ionization losses are similar. Thus, we treat the energy loss due to ionization as a bulk process that uniformly decelerates the ejecta, and consider spallation as the sole process that will affect the abundance pattern.
\end{enumerate}

We are interested in the early kinematics of the fast heavy particles ejected from an NSM.  Similar to supernova remnant evolution, we expect that the ejecta will initially be in free expansion, then sweep enough of the medium that shocks develop and the ejecta decelerates, until finally being stopped \citep[e.g.,][]{SNR}. Our focus is on the very first interactions while the ejecta energies are still high, above the thresholds for spallation.  This regime has not been well studied in the NSM case.  We select 1 yr after the explosion as the starting time for spallation, when the material is in the free expansion phase.
At this point, \textit{r}-process nucleosynthesis is finished, and the ejecta is expected to be moving with high velocity because there has been little interaction with the medium.

In these early phases during free expansion, we treat the initial particle interactions as scattering events rather than collective hydrodynamic motion.  We therefore describe the particle trajectories via the formalism for energy losses of fast particles moving through a medium.  Here, it is convenient to view the motion in the rest frame of the \textit{r}-process ejecta, with the medium being an incoming beam of interstellar composition (H and ${}^{4}{\rm He}$) moving at speed $v_E$.

\subsection{Propagation of the \textit{r}-Process Nuclei}
\label{subsec: propagation}

The transport equation for the \textit{r}-process ejecta can be written by adopting the expression for energetic particle propagation used in cosmic-ray studies \citep[e.g.,][]{CR1,CR2,Fields1994}:
\beq
\label{eq:fullprop}
\partial_t N_E  =  \partial_E (b_E N_E)  \ + q_E \
+ \mbox{escape} 
+ \mbox{advection} + \mbox{diffusion} \ .
\eeq
Here and throughout, $E$ denotes {\em kinetic energy per nucleon}, which depends only on the relative velocity between the projectile and target and, thus, is the same viewed from either frame. The instantaneous number of propagated particles per energy per nucleon at time $t$ is $N_E(E,t)=dN/dE$; thus, $N_E \ dE$ is the number of the propagated ejecta nuclei with kinetic energy in the range $(E,E+dE)$. The source function is $q_E=dN/dEdt$ and $b_E=-dE/dt$ is the rate of energy loss (per nucleon).
The number flux density is $\phi(E) = v(E) N_E$, where $v(E)=[1-(1+E/(m_pc^2))^{-2}]^{1/2}c$ is the velocity of the ejecta relative to the ISM and $m_p$ is the proton mass, such that for $v(E)=0.3c$, $E\sim45.29$ MeV. 

We simplify the propagation by neglecting the final terms \citep{Wang2018}:
\beq
 \label{eq:prop}
\partial_t N_E \approx  \partial_E (b_E N_E)  \ + q_E  \ .
\eeq
Here, we assume that the only important loss mechanism is the energy loss due to ionization and spallation reactions, and the escape term is thus ignored. This is for the following reasons.
(1) In the frame of the NSM ejecta, the incoming ISM particles are the projectiles, and so the energy losses they experience are due to interactions with the ejecta.  As discussed in Section~\ref{subsec: assumptions}, we are interested in the times after $\sim 1$ yr, 
when the ejecta is at least partially recombined, so that the energy losses are dominated by ionization losses in a neutral medium.  If the ejecta were still fully ionized, Coulomb losses are appropriate.  In practice, both of these loss mechanisms are due to the fast particle Coulomb fields, and both share the same scaling with density and particle speed, and have very similar magnitudes \citep[e.g.,][]{Mannheim1994}.  
(2) The pion creation is negligible in the MeV range. (3) Most nuclei in the ejecta are stable or radioactively decay over a much longer timescale than the energy loss timescale of $\sim 0.1$ Myr (see Appendix~\ref{sec: b}). We omit the advection and diffusion term in our model, as the electromagnetic interaction time is much shorter than the diffusion and advection times. In addition, spatial uniformity implies that the gradient-driven
advection and diffusion terms are zero. 
The thick-target model presented here neglects secondary particle effects.
By assuming that the \textit{r}-process nuclei lose energy continuously through the propagation, the effect of secondary light nuclei appears only via the elastic scattering energy loss term, and not as a proton/$\alpha$ particle source term. Secondary heavy particles are also not considered here. While these effects are not large, 
they would act only to boost spallation, and thus, our calculation can be considered a conservative estimate.

We assume the source functions for the projectile/initial \textit{r}-process nuclei $i$ with mass number $A_i$, charge $Z_i$ and number abundance $Y_i$ are
delta functions in time and in the projectile kinetic energy per nucleon $E_0$, i.e., the nuclei in the ejecta are all traveling in the same speed $v_E$ at time $t_0$ when ejected by the NSM:
\beq
\label{eq:source}
q_{i,\rm E}(E,t)= \frac{dN_i}{dt \, dE}=N_{i,0} \ \delta(E-E_0) \ \delta(t-t_0)=N'_{i,0}(E) \ \delta(t-t_0)
\eeq
where $N_{i,0}$ is the total particle number of \textit{r}-process element $i$, and where $E_0$ is the initial kinetic energy corresponding to initial velocity $v$.

We calculate a set of spallation reactions $i+j \rightarrow \ell + \cdots$ in which projectile $i$ and target $j$ nuclei give rise to products $\ell$.  A thick-target calculation described in Appendix~\ref{sec: equations} gives 
the product energy spectrum and energy-integrated production rate to be
\beqar
\label{eq:prod_spectrum}
q_{E,ij}^{\ell}(t) &=& n_{j} \ \sigma_{ij}^{\ell}(E) \ v(E) \ N_{i}(E,t),
\\
\label{eq:prod_number}
\frac{dN_{ij}^{\ell}}{dt} (t)&=& \int q_{E,ij}^{\ell}(t) \  dE 
\eeqar
where $n_j$ is the ISM number density of targets.
Here, $\sigma_{ij}^{\ell}(E)$ is the cross section for the production of nuclei ${\ell}$ by the reaction between ejecta nuclei $i$ and ISM nuclei $j$. 
As discussed in Appendix~\ref{sec: equations} and ~\ref{sec: b}, since $N_{i}(E,t)\propto 1/b_i(n_{\rm gas}, E)$ (Eq~\ref{eq:spectrum}), and $b \propto n_{\rm gas}$ (Eqs~\ref{eq:inelasticb} and \ref{eq:bi}), the gas density is exactly canceled in the numerator of the Eqs~\ref{eq:prod_spectrum} and \ref{eq:prod_number},
and thus, the final result is {\em independent of the gas density for the thick-target model}.

Thus, the number fraction of the total spallation-produced nuclei ${\ell}$ at time $t_f$ to the initial projectile $i$ at time $t_0$ is:
\beqar
\label{eq: fraction}
f_{i}^{\ell}&=& \sum_{j} f_{i,j}^{\ell}=\sum_{j} \frac{N_{ij}^{\ell}(t_f)}{N_{i,0}}=\sum_{j} y_{j} \int_{E_x(t_f)}^{E_0} \frac{\sigma_{ij}^{\ell}(E') \, v(E') \, dE'} {b_{i,E'}(n_{\rm gas})/n_{\rm gas}} \ , 
\eeqar
where $E_{0}$ is the initial kinetic energy per nucleon of the projectile nuclei $i$, which is the maximum kinetic energy of the nuclei. $E_x(t_f)$ is the kinetic energy per nucleon of the  projectile nuclei $i$ at time $t_f$ when the nuclei are no longer energetic enough to have spallation reactions.
The weighting $y_j=n_j/n_{\rm gas}$ is the fraction by number of ISM particles in the form of $j \in ({\rm H,He})$.

\paragraph{Order-of-magnitude estimate}

The fraction $f_{i,j}^{\ell}$ in Eq~\ref{eq: fraction} can also be expressed as an "optical depth" $\tau_{ij}^\ell$ :
\beqar
\label{eq: tau}
f_{i,j}^{\ell}
&=&\int_{E_x(t_f)}^{E_0} \frac{\sigma_{ij}^{\ell}(E')/m_{\rm u} dE'} {dE'/dX} = \int \kappa_{ij}^{\ell}dX= \tau_{ij}^\ell , 
\eeqar
where the spallation cross section sets in the "opacity" $\kappa^{\ell}_{ij} = \sigma^{\ell}_{ij}/m_{\rm u}$ and $m_{\rm u}$ is the nucleon mass. As the path-length of the nucleus $i$, i.e., the stopping distance or range, is $s_i=\int vdt = \int^E v(E) dE/b_{i,E}$, then $X=\rho_{\rm gas} s=n_{\rm gas}m_{\rm u} s$ is the grammage, which is independent of the medium density and has units  $\rm g cm^{-2}$. Our thick-target approximation assumes $\tau_{ij}^\ell < 1$, and in this limit, the optical depth $\tau_{ij}^\ell$ or $f_{i,j}^{\ell}$ represents the probability of a spallation reaction occurring, i.e., the fraction of nuclei $i$ that undergoes the reaction with the ISM nuclei $j$. 

Now we conduct an order-of-magnitude calculation to estimate the probability of the reaction between an ISM proton and nucleus $i$ with initial ejecta velocity $v$. With ionization loss as the dominant channel, we find that $b_{i,E}\sim b_{i, \rm ionic}\propto Z_i^2 n_{\rm gas}/(A_i v)$ from Eq~\ref{eq:bi}, and $X_i=\rho_{\rm gas}  \int^E v(E) dE/b_{i,E}\propto A_i v^4/Z_i^2$.  We thus have $f_{i,j}^{\ell}=A_i \sigma^{\ell}_{ij}/(m_{\rm u}Z_i^2)X_p$. For $v=0.4c$, the grammage for protons traveling through hydrogen is roughly $X_p\sim1.65\ \rm g/cm^2$. For the third \textit{r}-process peak, typical cross sections for p+$^{196}{\rm Pt}$ are $\sigma^{\ell}_{p,^{196}{\rm Pt}}\sim 1\ \rm barn$ \citep{Kusakabe2018}. Therefore, $f_{p,^{196}{\rm Pt}}^{\ell}\sim 0.032$, and for each of the dominant channels of the spallation reaction p+$^{196}{\rm Pt}$ approximately $\sim3\%$ of $^{196}{\rm Pt}$ is sandblasted. Several such spallation channels will be open, leading to a potential $\sim 10\%$ reduction in $^{196}{\rm Pt}$ and resulting increase in daughter species. This effect is large enough to be of interest, so we proceed with our detailed spallation calculations as described below and show the results in Section~\ref{sec: results}.

\subsection{Spallation Cross Sections of \textit{r}-Process Nuclei}
\label{subsec: spall}

\begin{figure}[H]
\centering
\includegraphics[scale=0.8]{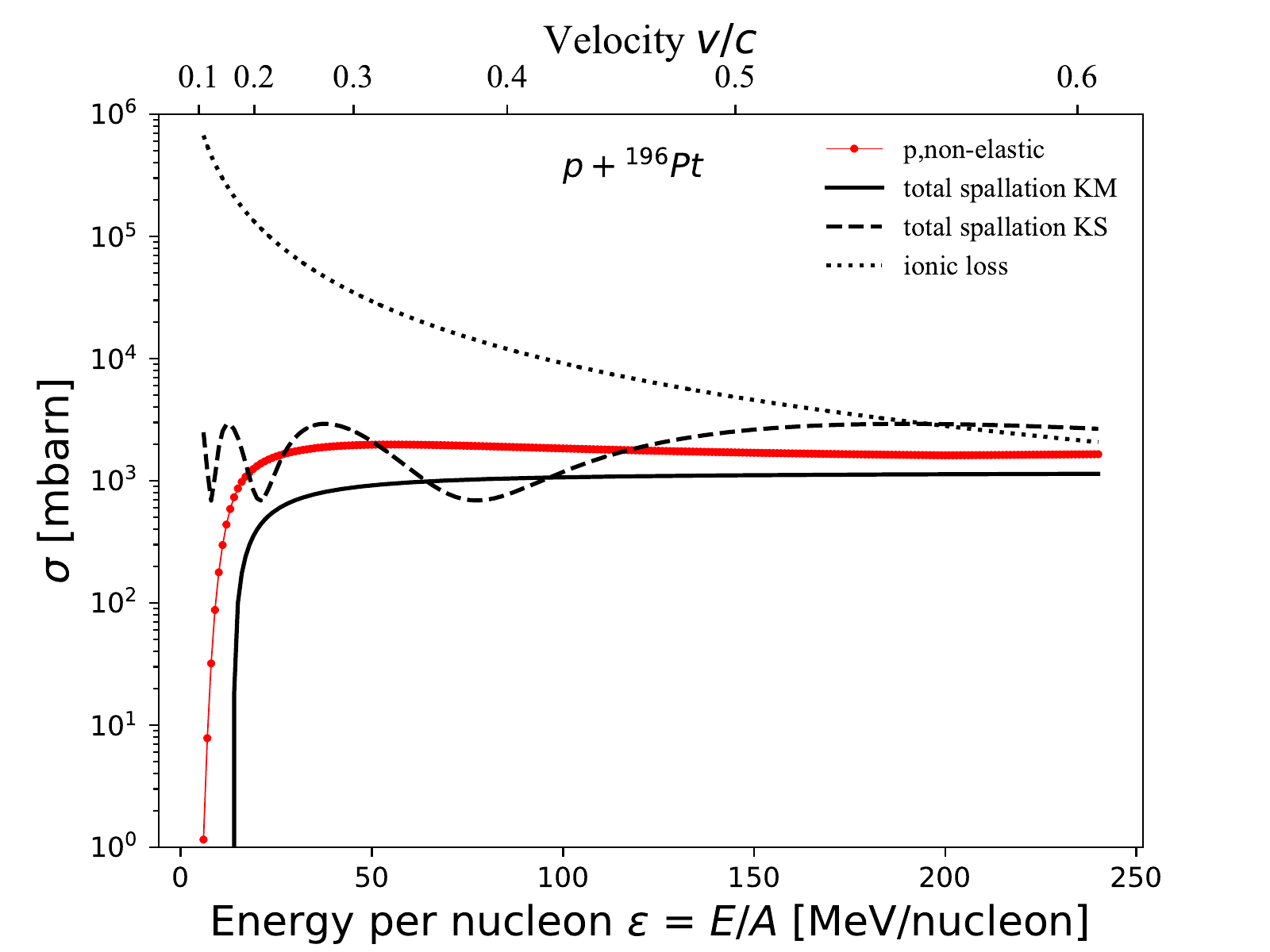}
\caption{
Total cross sections for the spallation reaction between $^{196}{\rm Pt}$ and a proton compared to the total ionization loss. The black dotted line is the ionic loss cross section, while the red dotted line is the total non-elastic/spallation cross section calculated using TALYS, the black solid line is the spallation cross section adopted in \citet[][hereafter KM]{Kusakabe2018}, and the black dashed line is the spallation cross section adopted in \citet[][hereafter KS]{Komiya2017}.
\label{fig:crosssection_total}}
\end{figure}

When propagating through the ISM, the \textit{r}-process nuclei ejected from a NSM lose energy mainly through the ionization of neutral hydrogen (or stopping power) in the MeV range, as discussed in in Section~\ref{subsec: propagation}. At the same time, these heavy nuclei also collide with ISM particles and fragment into lighter elements through spallation processes.  

In order to understand the production of new nuclei from the initial ejected \textit{r}-process material interacting with the ISM, we need to know the cross sections for nuclear spallation. 
Proton spallation reactions have been measured for a few target nuclides  \citep[e.g.,][]{spal11, spal10, spal9, spal8, spal7, spal6, spal5, spal4, spal3, spal2, spal1, spal0, spal00}.
However, there are little experimental data available for spallation reactions between a proton or $^{4}$He and a target nuclide that is heavier than iron, in the energy range smaller than $\lesssim100$ MeV. Therefore, in this paper, we adopt the theoretical spallation/inelastic cross sections from TALYS/1.9 \citep{TALYS,TALYS2}\footnote{https://tendl.web.psi.ch/tendl\_2019/tendl2019.html} with default nuclear inputs.

\paragraph{Total spallation cross section}
The ionization loss dominates over spallation loss, meaning that during propagation, the \textit{r}-process nuclei mainly lose energy through the electromagnetic interactions.
Figure~\ref{fig:crosssection_total} shows the comparisons between the total spallation and ionization loss cross sections for the isotope $^{196}{\rm Pt}$.  
The black solid line and dashed line are the spallation cross sections (proton reactions) adopted in \citet[][a semiempirical parameterization calculation from SPACS, \citet{Schmitt2014, Schmitt2016}]{Kusakabe2018} and \citet[][an empirical formula from \citet{Letaw1983}]{Komiya2017}, respectively. 
Compared with TALYS proton results shown in red, these simple formulae give cross sections of same order of magnitude, but 
the differences are not trivial. TALYS contains a variety of options for input nuclear physics such as nuclear level densities, gamma-strength functions, and optical potentials; variations in available inputs result in calculations that differ by at most 10 percent. However, different theory approaches can give larger variations, e.g., NONSMOKER calculations \citep{NS1,NS2,NS3}\footnote{https://nucastro.org/nonsmoker.html} result in cross-section values for the various spallation channels that can differ from TALYS by more than an order of magnitude. 
For this work, we calculate spallation effects using TALYS cross sections ($\sigma_{\rm TALYS}$) and with cross sections 10 times larger ($10\times \sigma_{\rm TALYS}$) to roughly account for these uncertainties.

\begin{figure}[H]
\centering
\includegraphics[scale=0.8]{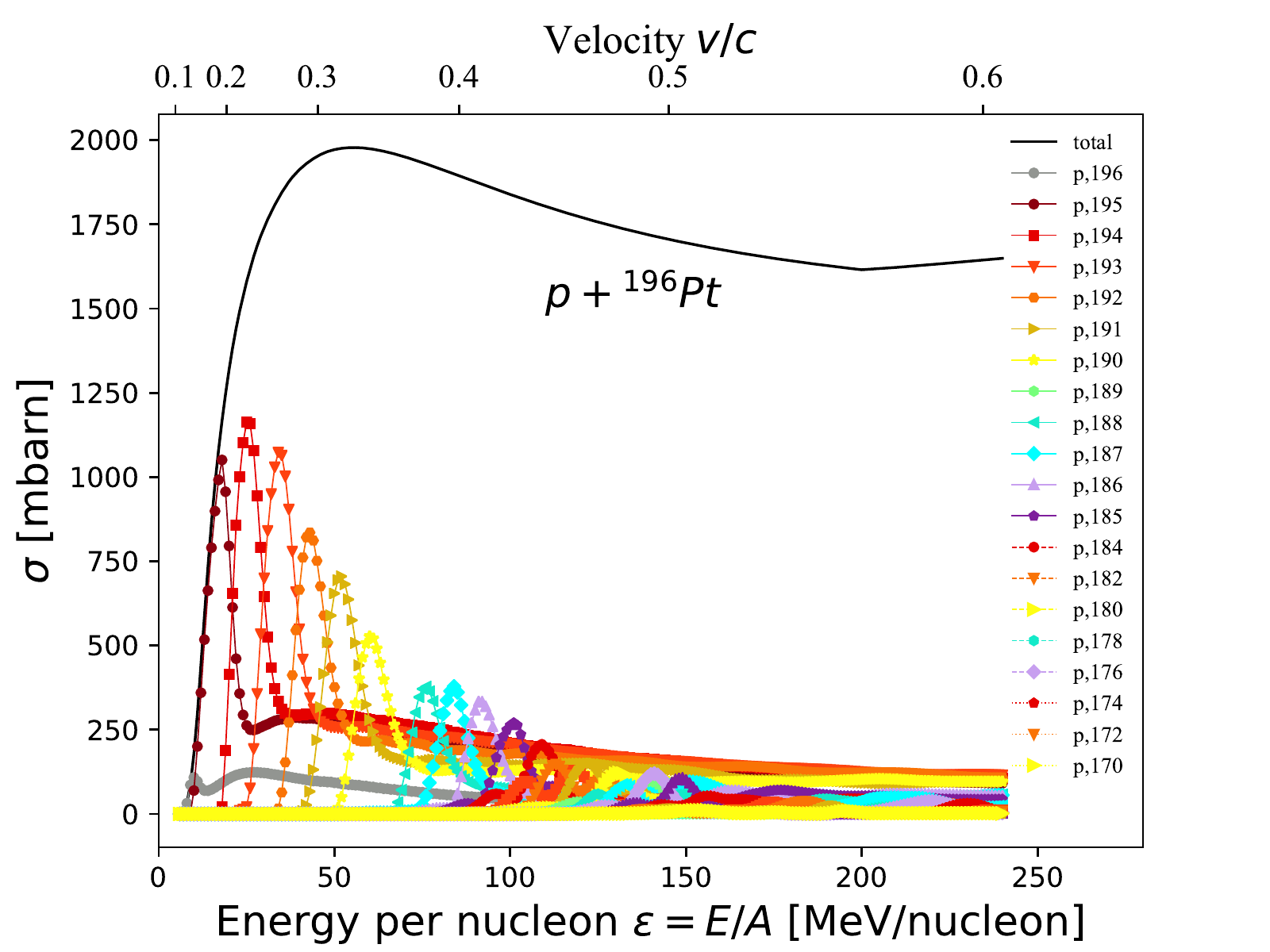}
\caption{
Individual cross sections for each spallation channel for the reaction between $^{196}{\rm Pt}$ and a proton, generated with TALYS. The black solid line is the total spallation cross section, while the colored lines show the cross sections for the individual channels $^{196}\mathrm{Pt}(p,x)A$, where $A$ is the mass number of the final nucleus after spallation. As the projectile energy per nucleon increases, the dominant spallation production channel moves from $A\sim195$ (dark red circles) to smaller mass numbers, with a wide range of product nuclei at the highest energies.
\label{fig:crosssection_channel}}
\end{figure}

\paragraph{Cross sections for each spallation channel}
Spallation reactions change a projectile nucleus to a new nucleus with a nearby but smaller mass number. 
Figure \ref{fig:crosssection_channel} shows the cross sections for each spallation channel $^{196}\mathrm{Pt}(p,x)$. The relevant energy range here is roughly $5-100$ MeV; moving from high projectile energy per nucleon to low within this range, the dominant creation channel shifts from producing $A=185$ to $A=195$ nuclei. Thus, we would expect spallation to shift the \textit{r}-process abundance pattern peaks to smaller mass numbers, and the results presented in Section~\ref{sec: results} confirm this expectation.

\subsection{Calculation of the Abundance Change due to Spallation}
\label{sec: abundance change}

From Figure~\ref{fig:crosssection_channel}, we can see that the number of nucleons that could be removed through spallation reactions depends on the projectile energy per nucleon, or the relative velocity of the projectile and target nuclei. With these spallation cross sections and Eq~\ref{eq: fraction}, we calculate the abundance change of the \textit{r}-process nuclei due to spallation, following the number conservation of the total ejecta particles shown in Eq~\ref{eq:f_prop}. 

From our order-of-magnitude estimate in Section~\ref{subsec: propagation}, we expect spallation to influence any individual abundance $Y_{i}$ by about 1-10\%. Neighboring nuclei tend to have similar spallation cross sections, varying by at most a factor of two. So for regions of the abundance pattern that are fairly flat, any particular mass number $A$ will be depopulated by spallation roughly at the same rate as it is repopulated by spallation of nuclei with higher $A$, and any significant rearrangement of the abundance pattern is unlikely. In the regions around the second ($110<A<140$) and third ($178<A<200$) \textit{r}-process peaks, there are steep abundance changes of an order of magnitude or more. A shift in abundance of 1-10\% here can produce a noticeable change to the peak shape, particularly on the lower-mass edge.
Thus, in this work,  we focus our spallation calculations exclusively on the two primary peak regions of the main \textit{r}-process pattern. 
We expect the spallation effects on other areas of the pattern to be smaller but show similar trends.

For $v<0.1c$ (corresponding to $E\lesssim4.73$ MeV/nucleon), the initial kinetic energy is too small to initiate spallation nuclear reactions,
so the \textit{r}-process ejecta needs to be faster than $0.1c$ for spallation reactions to occur. We consider initial ejecta speeds of 
$v/c=(0.2,0.3,0.4,0.5)$, corresponding to $E = (19,45,85,145)$ MeV/nucleon).
For $v=0.3c$, the spallation reactions will remove at most about 5 nucleons from the projectile nuclei (i.e., $A_i-5\leq A_{\ell}\leq A_i$); for $v=0.4c$, the spallation reactions will remove at most 10 nucleons from the projectile nuclei (i.e., $A_i-10\leq A_{\ell}\leq A_i$). 
Thus, we calculate the spallation effects, i.e., the abundance change of \textit{r}-process nuclei due to spallation, from the nearby $\sim10-25$ nuclei with heavier masses. 

For the third peak (second peak) of the \textit{r}-process abundance pattern, we check the abundances from $A=178$ to $A=200$ ($A=110$ to $A=140$), where the abundances peak at $A\sim196$ ($A\sim132$) and have a minimum at $A=185$ ($A=110$). As the abundances above $A=210$ ($A=150$) are negligible compared with the third peak (they are smaller and are in a much flatter shape compared with the second peak), we ignore the spallation effects for nuclei with $A>210$ ($150<A<178$). 

In addition, the ISM is mostly made up of protons ($\sim75\%$), and $\alpha$ particles ($^{4}$He) take up $25\%$ of the ISM total mass; thus, the number fractions of protons and $^{4}$He are $y_{p}\sim 6/7$, $y_{\alpha}\sim 1/7$. The cross sections for $\alpha$-$^{196}{\rm Pt}$ spallation reactions show similar trends to p-$^{196}{\rm Pt}$. 
Therefore, we include both proton and $\alpha$ particle spallation reactions with the initial \textit{r}-process ejecta nuclei to get the final spallation results.

For each nucleus $i$ interacting with ISM nucleus $j$ through spallation reaction $i+j\to \ell + \cdots$, we calculate $f_{i,j}^{\ell}$ ($A_{\ell}=[A_i-n,A_i-0]$). Because the particle number during propagation is conserved (eq.~\ref{eq:f_prop}), 
nucleus $i$ produces the same number of nucleus ${\ell}$; thus, the loss of nucleus $i$ during the propagation is at the same number as the production of all of the nuclei from the spallation reaction of nucleus $i$, i.e., $f_{i,\rm prop\ loss}=\sum\limits_{\ell} (y_{p}f_{i,p}^{\ell}+ y_{\alpha}f_{i,\alpha}^{\ell})$. The new abundance after spallation is therefore
\beqar
\label{eq:change}
Y_{i,\rm spallation} 
& = & Y_i(1-f_{i,\rm prop\ loss})+\sum\limits_{k} (y_{p}f_{k,p}^iY_k)+\sum\limits_{k} (y_{\alpha}f_{k,\alpha}^iY_k) \ \ .
\eeqar
To compare the new abundance pattern with the initial \textit{r}-process abundance pattern, we compute the spallation abundance change ratio by
\beqar
\label{eq:change_ratio}
F_{i,\rm {change}}=(Y_{i,\rm spallation}-Y_i)/Y_{i}. 
\eeqar

\section{Results}
\label{sec: results}

Spallation effects on \textit{r}-process nuclei ejected from an NSM depend both on the initial \textit{r}-process nucleosynthesis conditions and on the propagation process. The propagation process is affected by the velocity of the \textit{r}-process ejecta and spallation cross sections, while the astrophysical conditions and nuclear physics inputs determine the initial \textit{r}-process abundances generated from nucleosynthesis calculations. Thus both nuclear physics inputs and astrophysical conditions matter for our calculation, and spallation results could place constraints on these conditions in turn.

\paragraph{Nucleosynthesis calculation} In this work, we use the nuclear reaction network code Portable Routines for Integrated nucleoSynthesis Modeling \citep[PRISM;][]{Matt2016, Matt2017, Matt} to perform the \textit{r}-process nucleosynthesis calculations to obtain the abundance patterns for the initial \textit{r}-process nuclei ejected from an NSM. For our baseline nucleosynthesis calculation set, we begin with the nuclear masses from the 2016 Atomic Mass Evaluation \citep[AME2016;][]{AME} if available and FRDM2012 \citep{FRDM} otherwise. The Los Alamos National Laboratory (LANL) statistical Hauser-Feshbach code of \citet{HF} is used to calculate neutron-capture and neutron-induced fission rates for each nuclide. Branchings for $\beta$-delayed fission and $\beta$-delayed neutron emission are calculated using the LANL QRPA+HF code of \citet{LANL} as in \citet{Moller2019},  with strength functions and half-lives from \citet{Moller2003}.
Photodissociation is calculated using detailed balance, with one-neutron separation energies calculated directly from the combined FRDM2012 and AME2016 mass dataset. We first explore the case of a symmetric, two-fragment product distribution (symmetric split) for all fissioning nuclei. Finally, we take any values for spontaneous fission and $\beta$-decay rates from the Nubase2016 nuclear data evaluation \citep{Nubase} to replace any of our theory-based calculations. To gauge the influence of nuclear physics variations on our spallation results, we also adopt the $\beta$ decay rates of \citet{MT} and neutron-capture rates from NONSMOKER \citep{NS1,NS2,NS3}\footnote{https://nucastro.org/nonsmoker.html}, in addition to the baseline nuclear reaction rates.

We adopt two kinds of NSM trajectories to compare astrophysical conditions: cold dynamical ejecta \citep[e.g.,][]{Goriely2011, Matt} and a low entropy accretion disk wind which is parameterized similarly to conditions in \citet{McLaughlin2005, Surman2006, Just2015, Martin2015, Wanajo2014, Siegel2018}.

\subsection{Propagation Parameter Variations}
\label{subsec: propagationprocess}

For \textit{r}-process ejecta with a given initial abundance pattern, the spallation abundance change ratio is mainly dependent on the initial ejecta velocity and the spallation cross sections adopted. 
Here, we present our baseline simulation and explore the impact of variations in the initial ejecta velocity and in the spallation cross sections on the final abundances.

\begin{figure}[H]
\centering
\includegraphics[scale=0.4]{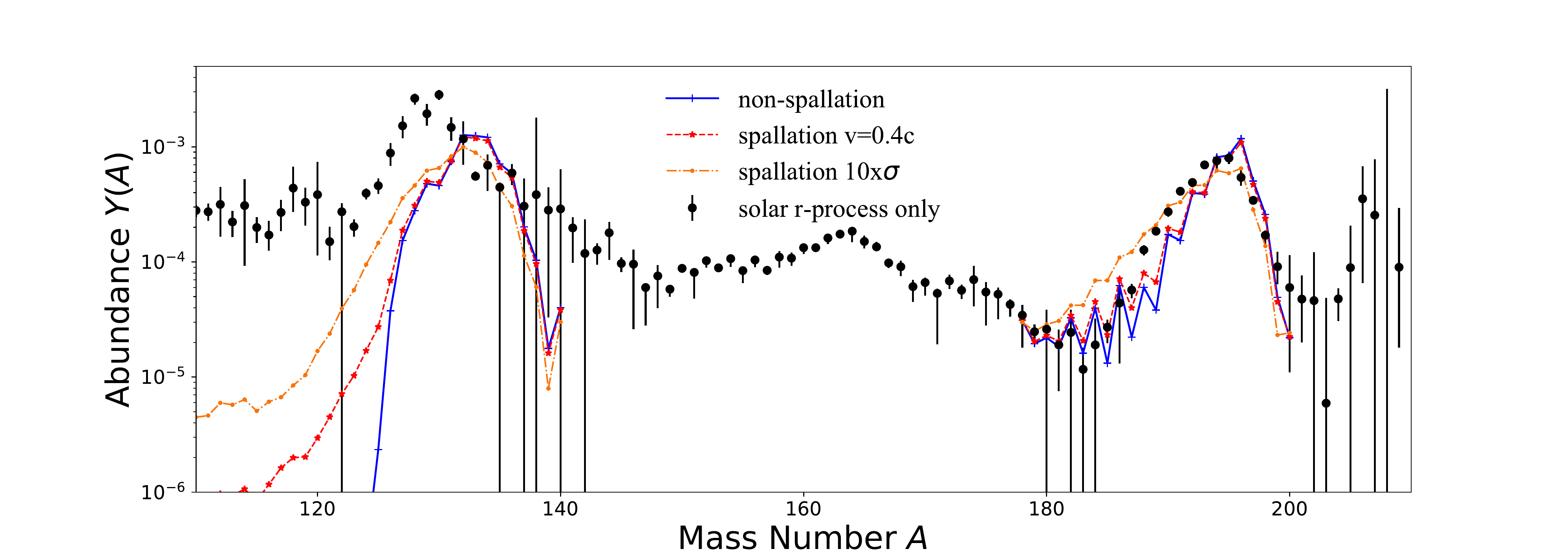}\\
\caption{Abundances in the second peak ($110<A<140$) and third peak ($180<A<200$) of the \textit{r}-process pattern produced with dynamical ejecta of a NSM with initial velocity 
$v=0.4c$ and two choices of spallation cross sections, $\sigma_{\rm TALYS}$ (red dashed line) and $10\times \sigma_{\rm TALYS}$ (orange dotted line). The initial \textit{r}-process abundance pattern from the PRISM simulation is shown in blue, and the black points are the solar \textit{r}-process residuals \citep{solar2007}. The solar data scales to the $^{195}$Pt abundance from the initial \textit{r}-process simulation.
\label{fig:spal_v}
}
\end{figure}

\paragraph{Baseline spallation calculation}
We first consider spallation effects on our baseline \textit{r}-process abundance pattern using the cold dynamical ejecta conditions described above. 
Figure~\ref{fig:spal_v} shows the resulting abundances of the second and third \textit{r}-process peaks before and after spallation calculations that assume an initial ejecta velocity of $0.4c$, and spallation cross sections from TALYS ($\sigma_{\rm TALYS}$) and $10\times \sigma_{\rm TALYS}$.
We can see that spallation moves the \textit{r}-process abundance pattern to lower mass numbers, toward the solar data, and smooths the shapes at the left side of the peaks while leaving the right side of the peaks largely unchanged. The spallation effect increases dramatically with an increased spallation cross section.

\begin{figure}[H]
\centering
\includegraphics[scale=1]{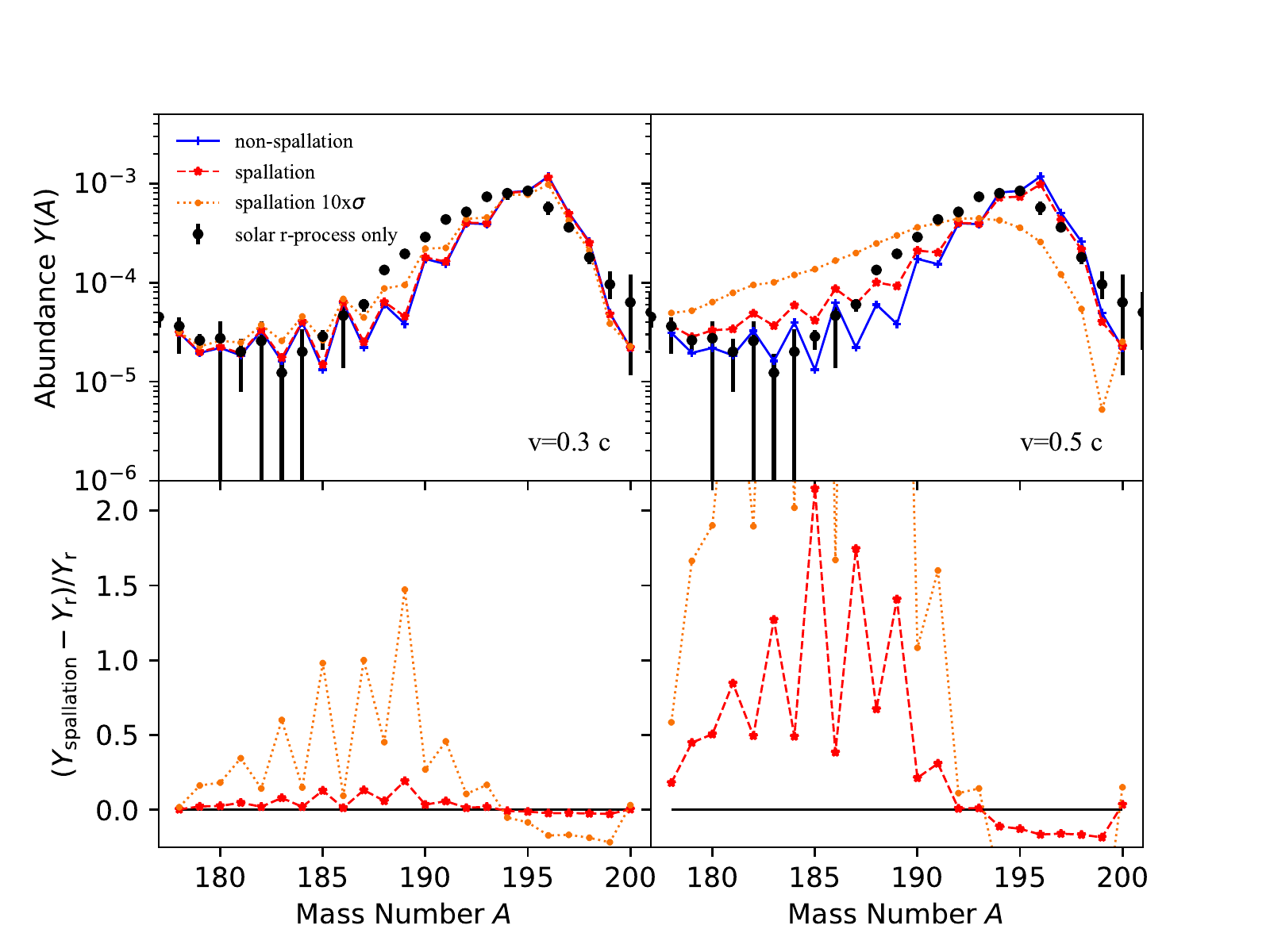}
\caption{Spallation effects in the third peak ($180<A<200$) region for the baseline dynamical ejecta simulation assuming initial ejecta velocities of $0.3c$ (left panels) and $0.5c$ (right panels). Upper panels: abundances before (blue lines) and after (red/orange lines) spallation, compared to solar data as in Figure~\ref{fig:spal_v}. Lower panels: the abundance change ratio due to spallation as defined in Eq~\ref{eq:change_ratio}.
\label{fig:spal_v3}
}
\end{figure}

\paragraph{Third \textit{r}-process abundance peak}
We next explore the effect of ejecta velocity on spallation in the $A\sim 195$ peak region. Figure~\ref{fig:spal_v3} 
shows the abundance pattern and the abundance change ratio due to spallation in the third peak region, starting from the same initial cold dynamical ejecta abundance pattern of Figure~\ref{fig:spal_v} and considering initial ejecta speeds of $0.3c$ and $0.5c$ for the calculation of the spallation effects.
Figures~\ref{fig:spal_v3} and \ref{fig:spal_v} together show that {\em the influence of spallation strongly depends on the velocity of the \textit{r}-process ejecta}, and the abundance pattern changes are nonnegligible for ejecta of $0.3c$ or faster: at $0.3c$, spallation brings, on average, $\sim 8\%$ of the abundance change and $\sim 60\%$ when the spallation cross section is increased by a factor of 10; at $0.5c$, the abundance change can be as high as a factor of 2 for some nuclei. The abundance shape is even flatter than the solar pattern with a velocity 
of $0.5c$ and adopted cross sections of $10\times \sigma_{\rm TALYS}$. This suggests that if the actual spallation cross sections are much higher than the predicted TALYS values, the bulk of the ejecta from an $\textit{r}$-process event cannot exceed the "speed limit" of about $\sim 0.4-0.5c$.

\begin{figure}[H]
\centering
\includegraphics[scale=1]{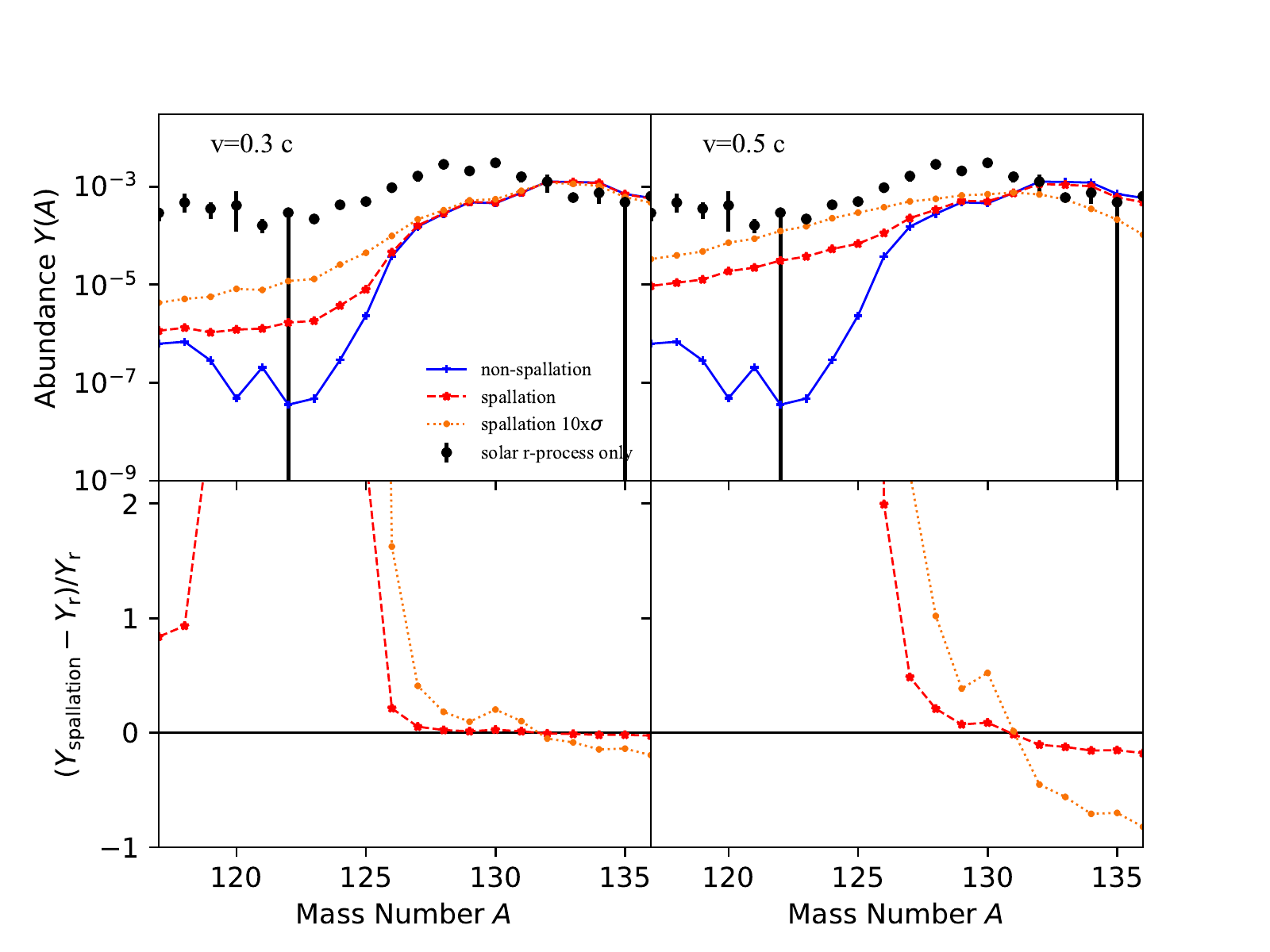}
\\
\caption{
Spallation effects in the second peak ($110<A<140$) region for the baseline dynamical ejecta simulation assuming initial ejecta velocities of $0.3c$ (left panels) and $0.5c$ (right panels). Upper panels: abundances before (blue lines) and after (red/orange lines) spallation; the solar data scales to the $^{132}$Xe abundance from the initial \textit{r}-process simulation. Lower panels: the abundance change ratio due to spallation as defined in Eq~\ref{eq:change_ratio}, scaled similarly to Figure~\ref{fig:spal_v3} to highlight the abundance changes in the second peak itself.
\label{fig:spal_v2}
}
\end{figure}

\paragraph{Second \textit{r}-process abundance peak}
Figure~\ref{fig:spal_v2} shows the abundance pattern and the abundance change ratio due to spallation in the second peak region, again for calculations starting from the initial dynamical ejecta abundance pattern of Figure~\ref{fig:spal_v} and considering initial ejecta speeds of $0.3c$ and $0.5c$ as in Figure~\ref{fig:spal_v3}. The initial abundances include only a main ($A>120$) \textit{r} process with no weak/limited \textit{r}-process component, so the left edge of the second peak is very sharp. Thus, the abundance changes due to spallation can be orders of magnitude larger here than for the third peak. Spallation helps to fill the gap with the solar data and smooths the shape of the peak. However, even at the highest initial ejecta speeds where spallation can largely fill in the region to the left of the peak, the fit to solar data remains poor.

It is important to note that while the third \textit{r}-process peak is produced only in the most robust neutron-rich environments, a wider variety of conditions can produce the second peak. Thus, the region to the left of the second peak is likely filled in with contributions from astrophysical trajectories with higher initial electron fractions $Y_{e}$ such as from the NSM accretion disk wind \citep[e.g.,][]{Erika}. While this higher $Y_{e}$ material may also undergo spallation in the ISM, the effects are likely smaller as the ejecta speeds are expected to be lower. The region to the left of the second peak can also be filled in with the products of heavy fissioning nuclei, an effect we explore in the next section.

\subsection{Initial \textit{r}-process Abundance Pattern Variations}
\label{subsec: initial}

In Section~\ref{subsec: propagationprocess}, we considered \textit{r}-process ejecta traveling through ISM with different (but still uniform) initial velocities and spallation cross section values, with a fixed initial abundance pattern. Here, we repeat the analysis of Section~\ref{subsec: propagationprocess} with different choices of nuclear physics and astrophysics conditions adopted for the nucleosynthesis, while keeping the initial velocity of the ejecta at $v=0.4c$.
Both astrophysical conditions and the choice of nuclear data adopted for the nucleosynthesis simulation affect the initial abundance pattern features. This leads to a variance in the potential influence of spallation. 

\paragraph{Third \textit{r}-process abundance peak}

\begin{figure}[H]
\centering
\includegraphics[scale=1]{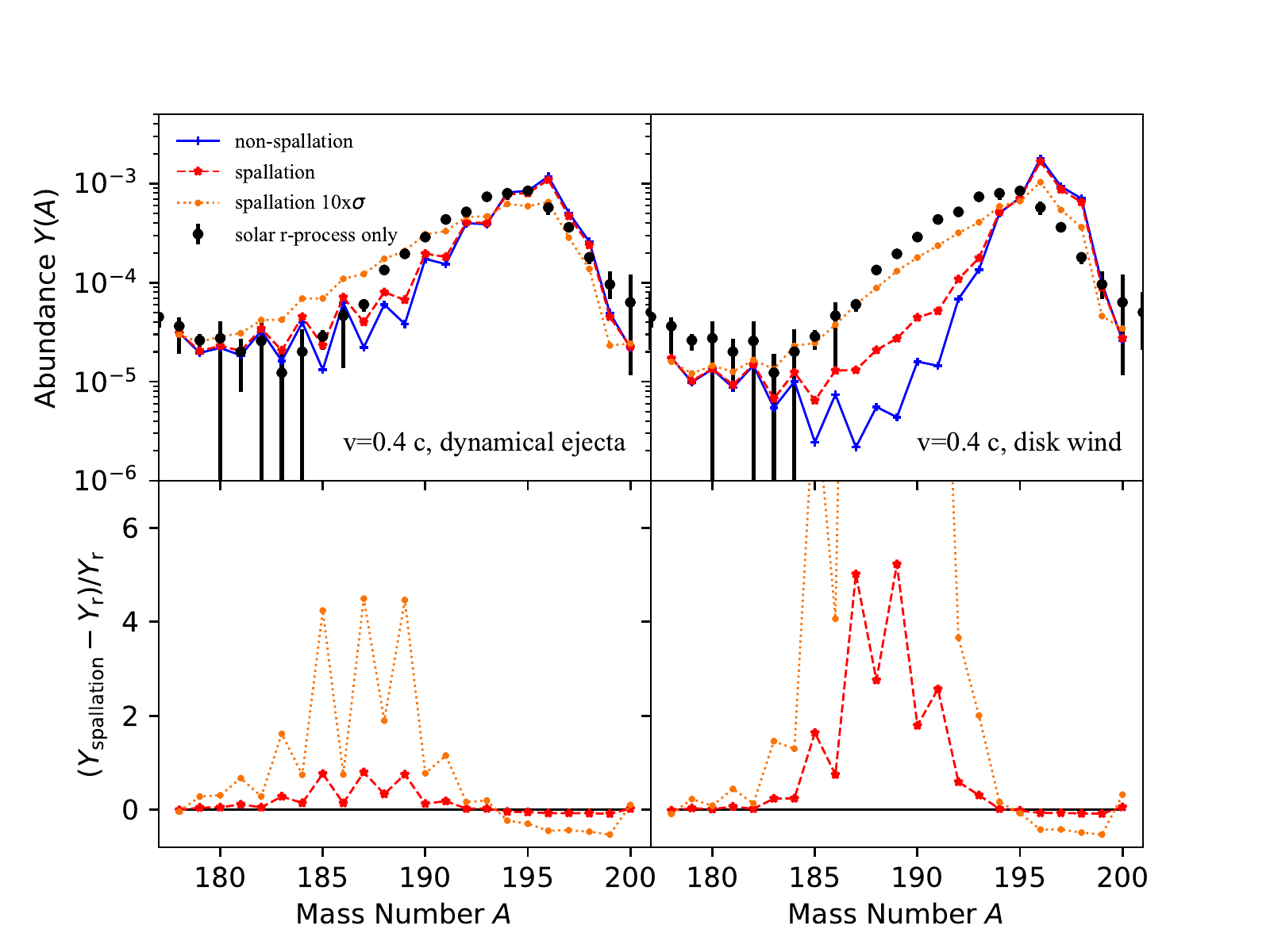}
\caption{
Spallation effects in the third peak ($180<A<200$) region for the baseline dynamical ejecta (left panels) and hot disk wind (right panels) simulations assuming initial ejecta velocity of $0.4c$. Upper panels: abundances before (blue lines) and after (red/orange lines) spallation, compared to solar data as in Figure~\ref{fig:spal_v}. Lower panels: the abundance change ratio due to spallation as defined in Eq~\ref{eq:change_ratio}.
\label{fig:spal_r3}
}
\end{figure}

Figure~\ref{fig:spal_r3} compares the abundance patterns and abundance change ratios after spallation for the baseline dynamical ejecta shown in Figure~\ref{fig:spal_v} (left) and hot disk wind conditions (right). Hotter \textit{r}-process freeze-out conditions are characterized by more late-time neutron capture, which produces abundance peaks that can be narrower than and offset from solar data, as shown in the blue lines of Figure~\ref{fig:spal_r3}. Though we recognize that $0.4c$ may be an unrealistically fast ejecta speed for disk winds, we still consider here whether spallation could possibly alleviate this mismatch. Indeed, with this high initial speed and with the larger ($10\times \sigma_{\rm TALYS}$) spallation cross sections, a very narrow, offset initial peak can be smoothed and shifted to produce a reasonable match with solar. In all cases, the effects of spallation are much larger with the sharper peak; the average positive spallation abundance change is $\sim200\%$ for the wind example versus $\sim50\%$ for the cold dynamical ejecta example. {\em Spallation effects are bigger for steeper abundance features}.

The \textit{r}-process proceeds through a region of the nuclear chart where the nuclear properties are highly uncertain \citep{Matt2016}. Different choices of nuclear data also yield different initial \textit{r}-process patterns. We repeat our spallation calculations starting with abundance patterns produced with different choices of nuclear data: $\beta$ decay rates from \citet{MT} and neutron capture rates calculated with NONSMOKER \citep{NS1,NS2,NS3}. Both sets of rates act to broaden the third peak and move the peak position towards solar, resulting in flatter abundance shapes and smaller abundance changes due to spallation compared to those in Figure~\ref{fig:spal_r3}.

\paragraph{Second \textit{r}-process abundance peak}

{Fission may play an important role in shaping the second peak of the \textit{r}-process abundance pattern. The simple symmetric split adopted for fission products in the baseline calculation deposits material directly into the $N=82$ region, resulting in the strong and concentrated initial $A\sim130$ peak shown in Figure~\ref{fig:spal_v2}. However, fission yields are expected to be distributed in both proton number and mass number. The widths of fission fragment distributions will influence the height and width of the second \textit{r}-process peak, resulting in a range of potential spallation effects. 

A variety of theoretical fission yield prescriptions have been implemented in \textit{r}-process calculations, e.g., \citet{KT,GEF,Goriely2013,Shibagaki2016,Mumpower2019, Nicole1}.
Here, we compare our baseline simulation to an identical simulation adopting GEF 2016 \citep[version GEF-2016-V1-2,][]{GEF} yields from \citet{Nicole}.
GEF 2016 predicts a global trend of transition from asymmetric toward symmetric yields along most isotopic chains, with wide distributions of daughter products. As shown in \citet{Nicole}, GEF 2016 yield distributions result in a shallower second peak and deposition into the region to the left of the peak. Thus, the influence of spallation is expected to be reduced compared to the baseline calculation.}

\begin{figure}[H]
\centering
\includegraphics[scale=1]{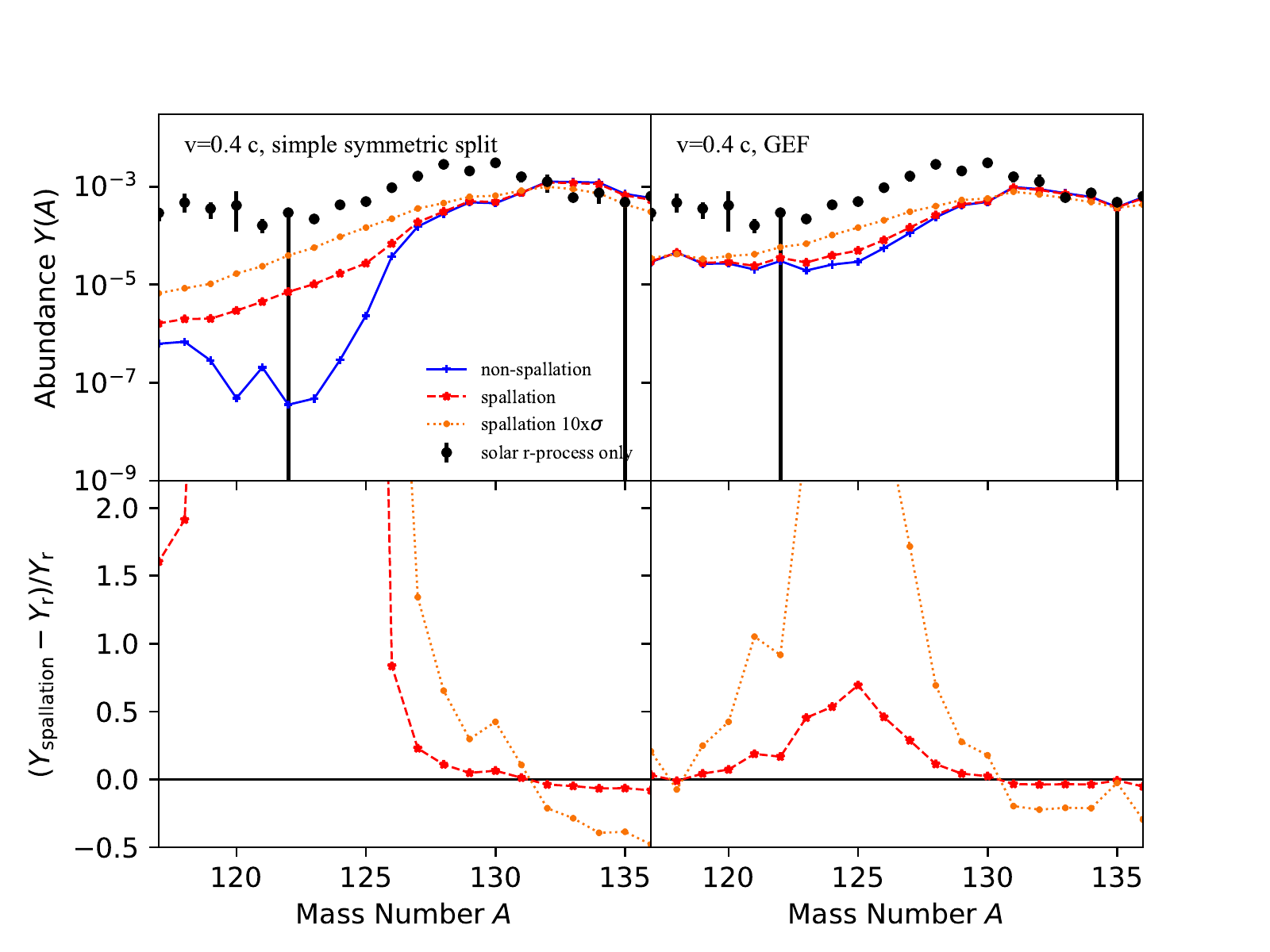}
\caption{
Spallation effects in second peak ($110<A<140$) region for the baseline dynamical ejecta simulation assuming initial ejecta velocity of $0.4c$, with two choices of fission yields (left panels: simple symmetric split; right panels: GEF 2016 fission yields \citep{GEF}). Upper panels: abundances before (blue lines) and after (red/orange lines) spallation, the solar data scales to the $^{132}$Xe abundance from the initial \textit{r}-process simulation. Lower panels: the abundance change ratio due to spallation as defined in Eq~\ref{eq:change_ratio}, scaled similarly to Figure~\ref{fig:spal_v2} to highlight the abundance changes in the second peak itself.
\label{fig:spal_r2}
}
\end{figure}

Figure~\ref{fig:spal_r2} shows the abundance pattern and abundance change ratio after spallation for the baseline cold dynamical ejecta conditions with two choices of fission yields (left panels: simple symmetric split, as in Figure~\ref{fig:spal_v}; and right panels: GEF 2016 fission yields). 
We can see that, compared to simple symmetric split, GEF 2016 fission yields fill the huge gap on the left side of the second peak, bringing a flatter abundance pattern and accordingly smaller spallation effect (GEF 2016: $\sim20\%$ on average; simple symmetric split: $\sim10,000\%$ on average).
We also tested Kodama $\&$ Takahashi fission yields \citep{KT} and found the abundance pattern to be even flatter than solar data, and spallation further enhances the disagreement.  

Moreover, we tested spallation on the second peak with different astrophysical conditions, including those where fission plays a minor role if any. In general, we find that the effects of spallation on the second peak can vary by $\sim2-3$ orders of magnitude when different initial conditions are adopted. Thus, the variation in spallation abundance changes due to initial conditions is larger than the variation ($\lesssim1$ order of magnitude) due to different spallation cross sections. Similar to the third peak, the effects of spallation are the largest where the ``cliff" to the left of the peak is sharpest.

\subsection{A Full NSM Simulation}
\label{subsec: simulation}

All of our analysis described previously considers only individual astrophysical trajectories with set electron fractions and initial velocities. {In reality, we expect \textit{r}-process material to be ejected from a neuron star merger event with a distribution of velocities and other hydrodynamical conditions. Here, we post-process tracer particle trajectories from an NSM simulation to examine how the range of ejecta velocities, and thus the range of resulting spallation effects, influences the nucleosynthetic outcome of an NSM event.}

\begin{figure}[H]
\centering
\includegraphics[scale=0.85]{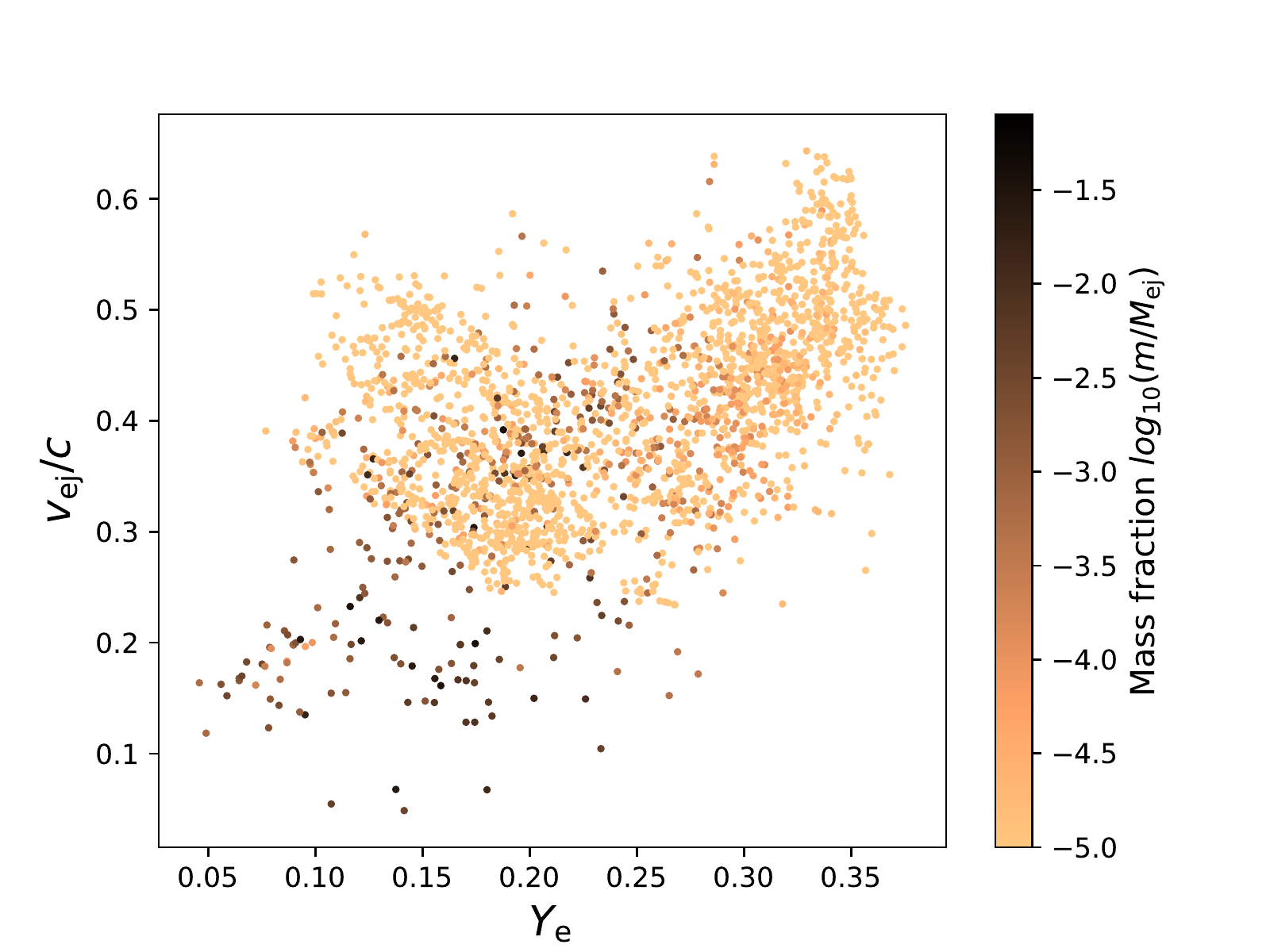}
\caption{Distribution of the electron fraction $Y_e$ and ejected velocity $v_{\rm ej}$ for the SFHO-M1.35 model from the NSM simulation of \citet{Bovard}. Different colors mark tracers with different mass fractions $m/M_{ej}$. The total ejected mass is $M_{ej}=3.53\times10^{-3}\msol$. The simulation gives 2253 tracers with mass averaged $\avg{Y_e}=0.16$ and $\avg{v_{\rm ej}}=0.26c$.  
}
\label{fig: velocity}
\end{figure}

{We choose the SFHO-M1.35 model from the relativistic NSM simulations of \citet{Bovard} as our example case. This model includes 2253 tracers with a mass-averaged velocity of 0.26$c$, consistent with other simulations \citep[e.g.,][]{Korobkin2012, Bauswein2013, Hotokezaka2013} of NSM dynamical ejecta and the kilonova observations discussed in Section 1. The velocities of the individual tracers at the onset of nucleosynthesis are shown in Figure~\ref{fig: velocity}. While most of the mass is ejected at velocities of $\sim$0.1-0.3$c$, a wide range of velocities up to 0.6$c$ are obtained.}

\begin{figure}[H]
\centering
\includegraphics[scale=0.4]{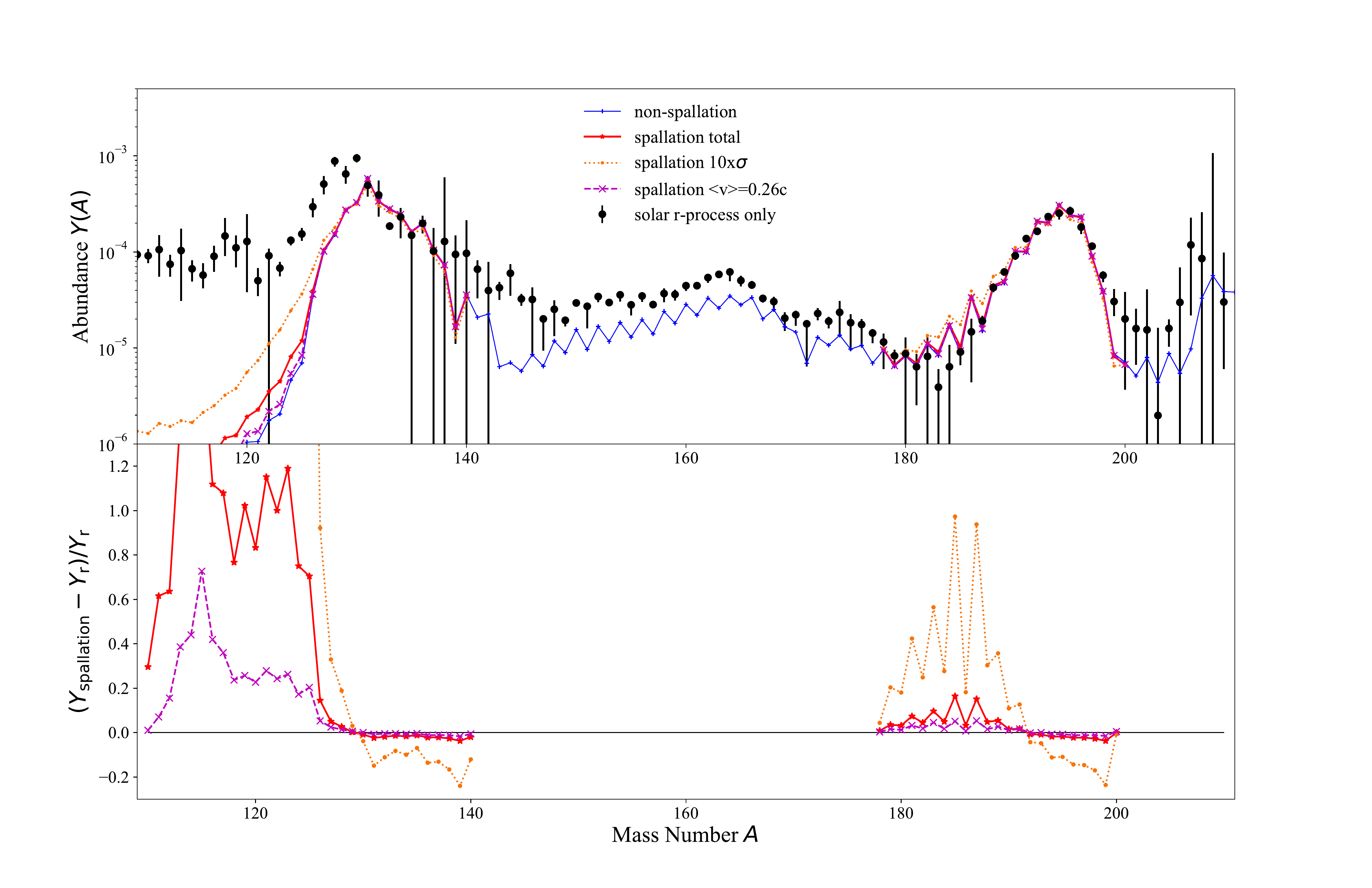}
\caption{
Spallation effects on the second peak ($110 < A < 140$) and third peak ($180 < A < 200$) regions of the \textit{r}-process abundance pattern produced with baseline PRISM calculation for a full NSM simulation \citep{Bovard} with two choices of spallation cross sections, $\sigma_{\rm TALYS}$ (red/purple lines) and $10\times \sigma_{\rm TALYS}$ (orange dotted line). The initial \textit{r}-process abundance pattern from the PRISM simulation is shown in blue and the black points are the solar \textit{r}-process residuals \citep{solar2007}. The solar data scales to the total abundance of the initial third abundance peak ($A=180-200$) of the \textit{r}-process simulation before spallation.\\
Upper panel: abundances before (blue lines) and after (red/orange/purple lines) spallation. Red and orange lines are the mass-summed abundance patterns after spallation from each tracer, the magenta line is the abundance pattern after spallation with single mass-averaged velocity $v\sim0.26c$.
Lower panel: the abundance change ratio due to spallation as defined in Eq~\ref{eq:change_ratio}, scaled similarly to Figure~\ref{fig:spal_v3} to highlight the abundance changes in the second and third peaks themselves.
\label{fig: total_sfho}
}
\end{figure}

{We calculate the element synthesis for each tracer trajectory using PRISM as described in Section 3. We add up the resulting $r$-process abundances of each tracer based on its mass to obtain the total 
(pre-spallation) abundance pattern for the merger event, shown by the blue solid line of  Figure~\ref{fig: total_sfho}. Then, we proceed with our spallation calculations, as in Section 2.2 in the following two ways for comparison: (1) We perform the spallation calculation for each tracer abundance pattern assuming the initial velocity for that tracer extracted from the NSM simulation. We then calculate the mass weighted sum of all of the post-spallation abundance patterns, with the result shown by the red solid line in  Figure~\ref{fig: total_sfho}. We also increase the spallation cross sections by a factor of 10 and repeat the procedure, obtaining the abundance pattern shown by the orange dotted line. (2) We perform a single spallation calculation on the total abundance pattern (blue line) assuming a mass-averaged initial velocity of $0.26c$. The result is shown in the purple line of  Figure~\ref{fig: total_sfho}.}

\begin{figure}[H]
\centering
\includegraphics[scale=0.4]{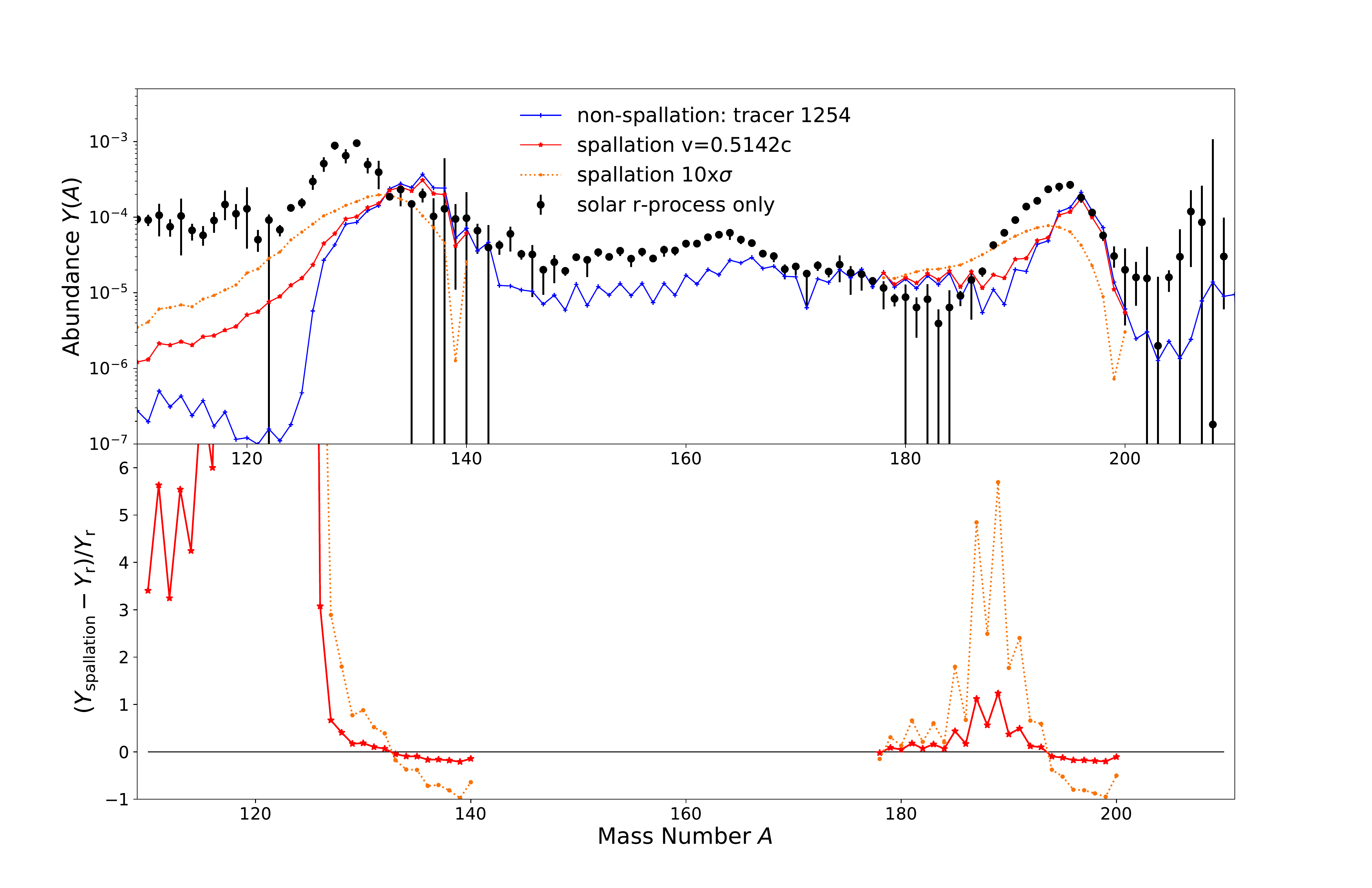}
\caption{
Spallation effects on the second peak ($110 < A < 140$) and third peak ($180 < A < 200$) regions with a baseline PRISM calculation for the high-speed tracer No.1254 from the NSM simulation with initial ejecta velocity of $v\sim0.512c$ \citep{Bovard}.
Upper panel: abundances before (blue lines) and after (red/orange lines) spallation for the tracer No.1254, compared to solar data, as in Figure~\ref{fig: total_sfho}. Lower panel: the abundance change ratio due to spallation as defined in Eq~\ref{eq:change_ratio}, scaled similarly to Figure~\ref{fig: total_sfho} to highlight the abundance changes in the second and third peaks themselves.
\label{fig: sfho_1254}
}
\end{figure}

Comparing the red and orange lines with the blue line in Figure~\ref{fig: total_sfho}, we find that spallation produces noticeable changes to the abundance pattern of a full NSM simulation and that the effect increases with an increased spallation cross section. The positive abundance change due to spallation is around $10\%$ on average for the third peak and $100\%$ on average for the second peak. 
Comparing the red line and purple line, we can see that the estimated effects of spallation on a single combined trajectory are smaller than those of the full calculation. This is mainly due to the contributions of the tracers with the highest speeds, especially for those beyond $0.5c$. For example, Figure~\ref{fig: sfho_1254} shows the abundance pattern and abundance change ratio after spallation for tracer No.1254 with $v=0.5142c$. The blue line is the initial \textit{r}-process abundance pattern, which has steeper shapes at both the second and third peaks compared to the combined trajectory. The high velocity and the steepness of the abundance peaks together result in larger abundance changes due to spallation. As Figure~\ref{fig: total_sfho} shows, the faster components of the ejecta ($v\ge0.5c$) do end up influencing the overall yields of an NSM event even though their net mass is small ($\sim0.5\%$ of the total ejecta mass).

{It is important to point out that the tracers considered in Figure~\ref{fig: total_sfho} are associated with only the dynamical outflow of the NSM event. In this simulation, the material is very neutron rich, with a mass-averaged electron fraction of $<Y_{e}>=0.16$, and produces primarily main \textit{r}-process nuclei ($A>120$). Recent simulations of dynamical ejecta by other groups  \citep{Goriely2015,Radice18} show a broader range of initial electron fractions and corresponding production of lighter \textit{r}-process nuclei. Additional mass outflows from, e.g., the resulting accretion disk, are expected. These outflows can produce weak ($70<A<120$) and possibly main \textit{r}-process nuclei that contribute to the overall NSM nucleosynthetic yields \citep{Wanajo2014,Martin2015, Thielemann2017}. Thus, the left side of the $A\sim 130$ peak is potentially formed from a variety of nucleosynthetic contributions where spallation is a subdominant effect.
Still, we expect that for NSM outflows, the full distribution of mass ejecta velocities should be considered when estimating the effects of spallation, since assuming a constant averaged velocity can result in an underestimate, such as that illustrated in Figure~\ref{fig: total_sfho}.}

\subsection{Spallation on Elemental Pattern/Elemental Ratio}
\label{subsec: element}

So far, we have focused our attention on the influence of spallation on the isotopic \textit{r}-process pattern. However, isotopic \textit{r}-process abundances are limited to solar-system data, while elemental \textit{r}-process abundances are available for a growing number of stars \citep{Sneden2008, Roederer2014}.

\begin{figure}[H]
\centering
\includegraphics[scale=0.9]{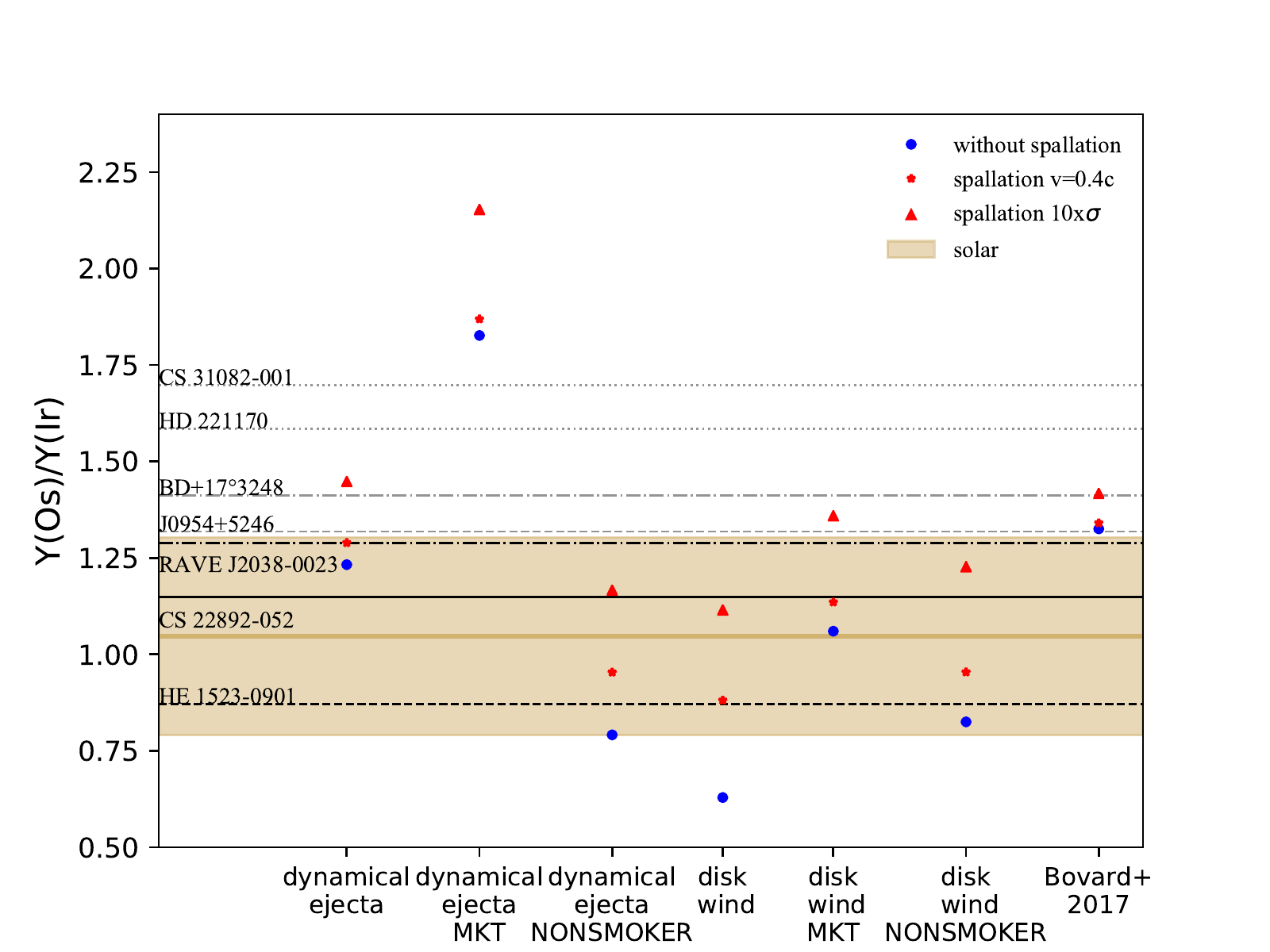}
\caption{
Elemental abundance ratios of Os over Ir (Y(Os)/Y(Ir)) for the dynamical ejecta and hot disk wind simulations assuming initial ejecta velocity of $0.4c$ (except for the NSM simulation \citep{Bovard}, which gives the mass-summed results of tracers with various velocities), with three choices of adopted nuclear data (PRISM baseline calculation, $\beta$ decay rates from \citet{MT} (MKT), and neutron capture rates from NONSMOKER \citep{NS1,NS2,NS3}\footnote{https://nucastro.org/nonsmoker.html}). The ratios are compared with solar data \citep[brown line: solar value, brown region: uncertainty range;][]{Sneden2008} and stellar observations of the \textit{r}-process enhanced stars (lines):  CS 31081-991 \citep{Hill2002}, HD221170 \citep{Ivans2006}, BD+17$^{o}$3248 \citep{Cowan2002}, CS 22892-052 \citep{Sneden2003}, HE1523-0901 \citep{Frebel2007}, J0954+5246 \citep{Erika2018}, RAVE J2038-0023 \citep{Placco2017}. 
\label{fig: ab_zr}
}
\end{figure}

Therefore, we also test the effects of spallation on the \textit{r}-process elemental pattern (abundance $Y$ versus charge number $Z$) and compare with the stellar observations. For this analysis, we consider the well-measured third peak elements Os and Ir from \textit{r}-process enhanced stars; for example, CS 31081-991 \citep{Hill2002}, HD221170 \citep{Ivans2006}, BD+17$^{o}$3248 \citep{Cowan2002}, CS 22892-052 \citep{Sneden2003}, HE1523-0901 \citep{Frebel2007}, J0954+5246 \citep{Erika2018}, RAVE J2038-0023 \citep{Placco2017}. Os/Ir ratios (Y(Os)/Y(Ir)) from solar data and these stellar observations have a wide range (from $\sim 0.8$ to 1.7). We compare the Os/Ir ratio from different \textit{r}-process nucleosynthesis simulations (blue circles) to the observation data (lines) in Figure~\ref{fig: ab_zr}. We find that most simulations give the Os/Ir ratio within the solar error bar and arevconsistent with the stellar data. Spallation can adjust this ratio, and we see that although the effect (red stars and triangles) is relatively small, in several cases it moves the Os/Ir ratio more comfortably within the observation range. The one outlier is the case of the cold dynamical ejecta conditions calculated with $\beta$ decay rates from \citet{MT}, where the simulation ratio is already bigger than the observation range. Therefore, for this case, spallation acts in the opposite direction, away from observational data, potentially putting constraints on the ejecta velocity, spallation cross sections, or nuclear data.

\subsection{Spallation Cross-section Sensitivity Study}
\label{subsec: sensitivity}

There is little experimental data available for heavy element spallation reactions, especially at the energies smaller than $100$ MeV of interest here, and different theoretical estimates can vary by as much as an order of magnitude. As shown in the results throughout this work, the effects of spallation are sensitive to the spallation cross sections adopted. Thus, we perform a sensitivity study to identify the nuclei whose spallation cross-section adjustments would bring the largest changes to the abundance pattern. We focus on the third peak region, since the second peak shapes are affected by initial nucleosynthesis inputs like fission yields more significantly than spallation cross sections (see Section~\ref{subsec: initial}).

For our sensitivity study, we start with the third peak abundance pattern of the baseline \textit{r}-process simulation (dynamical ejecta with $v=0.4c$). We increase the spallation cross section by a factor of 10 for one nucleus at a time, while keeping the other parameters and the cross sections for all other nuclei unchanged, and repeat the spallation calculation (single-increased scenario). We report the results for the most impactful cross sections in Table~\ref{tab:sensitivity}. Here, we consider the nuclei with increased abundances (i.e., positive change ratios) after spallation, and we average their abundance change ratios to obtain the values reported in the table. 

We list the top 10 nuclei that result in the largest average positive abundance change ratios.
For comparison, Table~\ref{tab:sensitivity} also shows the average positive abundance change ratios for the calculation with the original TALYS calculated spallation cross-section values (original scenario) and the case where TALYS cross sections for all nuclei are increased by a factor of 10 (all-increased scenario). The effects of spallation for the single-increased scenario for $ ^{198}$Pt, $ ^{197}$Au, $ ^{196}$Pt, and $ ^{195}$Pt are $\sim 80\%$ of the size of the full effects of the all-increased scenario, suggesting that spallation of the third \textit{r}-process peak is the most sensitive to these four cross sections. 

\begin{table*}
\centering

\begin{tabular}{|c|c|c|}
\hline\hline
\multicolumn{2}{|c|}{Adopted Cross Sections} & Spallation Effect \\
\multicolumn{2}{|c|}{$\sigma_{\rm spallation}$}  & [percentage] \\
\hline
\multicolumn{2}{|c|}{TALYS value (1$\times$)} & 23.99 \\
\multicolumn{2}{|c|}{(original)}  & \\
\hline
\multicolumn{2}{|c|}{10$\times$ for all nuclei} & 138.19 \\
\multicolumn{2}{|c|}{(all-increased)}  & \\
\hline
\multirow{10}*{10$\times$ for } & &  \\
\multirow{10}*{each individual nucleus} & $ ^{198}$Pt & 114.18 \\
\cline{2-3}
& $ ^{197}$Au & 112.54 \\
\cline{2-3}
& $ ^{196}$Pt & 108.91 \\
\cline{2-3}
& $ ^{195}$Pt &106.98 \\
\cline{2-3}
& $ ^{194}$Pt & 91.78 \\
\cline{2-3}
& $ ^{193}$Ir & 64.00 \\
\cline{2-3}
& $ ^{192}$Os & 45.49 \\
\cline{2-3}
& $ ^{191}$Ir & 54.67 \\
\cline{2-3}
& $ ^{190}$Os & 46.22 \\
\cline{2-3}
& $ ^{189}$Os & 33.55 \\
\hline\hline
\end{tabular}\\
\caption{Spallation cross section sensitivity study starting with the baseline dynamical ejecta simulation with $v=0.4c$ and focused on the third peak of \textit{r}-process abundance pattern. 
The table lists the average positive abundance change ratio (``spallation effect") resulting from the original, all-increased, and single-increased scenarios as described in the text. For the single-increases scenario, nuclei with the top 10 most impactful spallation cross sections are shown.
\label{tab:sensitivity}
}
\end{table*}

\section{Discussions and Conclusions}
\label{sec: discussion}

In this paper, we report on the construction and results of a thick-target spallation model to test the effects of spallation on the isotopic abundance pattern of nuclei ejected from NSM events. We find that spallation can result in nonnegligible changes to relative abundances in the $A\sim 130$ and $A\sim 195$ \textit{r}-process peak regions of material ejected with speeds of $0.3c$ and above.

The effects of spallation are to move the abundance pattern toward lower mass numbers and smooth the slope at the left side of the peaks. 
This effect depends both on the initial \textit{r}-process nucleosynthesis conditions and on the propagation process. 
The abundance pattern before spallation is set by the initial astrophysical conditions and the adopted nuclear data; we find that spallation can produce larger changes to abundance peaks that are initially sharper or steeper. Spallation occurs as the ejecta propagates through the ISM; thus, faster initial ejecta speeds result in more significant abundance changes. We find that spallation can partially or fully alleviate the mismatch of the second and third \textit{r}-process abundance peaks compared to solar data. {\em Thus the effects of spallation are non-negligible and need to be considered in addition to other factors for shaping the \textit{r}-process abundance pattern, especially for the trajectories having steeper abundance peaks and with initial speeds over $0.3c$.}

Moreover, the spallation effects can potentially put constraints on the initial \textit{r}-process nucleosynthesis conditions and propagation process in turn. The \textit{r}-process abundance pattern after spallation should not be flatter than the solar data. 
Our tests show that such a mismatch can occur for some choices of input nuclear data and for high ejecta speeds, and they furthermore suggest that if spallation cross sections are a factor $\sim 10$ above their theoretical estimates, the bulk \textit{r}-process ejecta has a `speed limit' of about $0.4-0.6c$, based on its initial abundance shape.

There still remains space to improve our model. 
A more complete treatment of the particle phase space evolution could take a Boltzmann-like approach to follow the transition from individual particle scattering interactions to the development of shocks.  We ignore diffusion here, but in reality, the NSM environment may have strong magnetic fields that could be carried with the ejecta.  The motions of both the \textit{r}-process nuclei and the ISM particles may be quite complex if magnetic fields remain strong. In this scenario, the \textit{r}-process nuclei may have more time to undergo spallation reactions, changing the spallation/ionization reaction ratio and thus increasing the spallation effects in a longer timescale.  Theoretical work would also benefit from additional studies of spallation reactions and \textit{r}-process nucleosynthesis with different mass models. 

As the spallation cross sections we have adopted are from theoretical calculations, {\em our findings call for new experiments}.  Measurements of spallation reactions of \textit{r}-process heavy nuclei at relevant energies can pin down the true behavior of the cross sections and branching ratios.  Such data will establish the importance of spallation in altering and diagnosing NSM \textit{r}-process abundances. As seen in Figure \ref{fig:crosssection_channel}, the important energy range is $\sim5-100$ MeV, and the most important projectiles are protons, but $\alpha$ particle interactions deserve study as well. The most critical spallation targets are for the $A \sim 196$ nuclides listed in Table~\ref{tab:sensitivity}.  Because most spallation reactions occur long after the merger, they involve heavy nuclei at or near stability and, thus, sidestep the well-known challenges of studying the highly unstable nuclei essential to the \textit{r}-process synthesis itself. Thus, these measurements could be
within reach for appropriate facilities such as FRIB, FAIR, and RIKEN .

\acknowledgments
We are very grateful to Luke Bovard for providing us with trajectories from his neutron star merger simulations. 
We are pleased to thank George Fuller, Wick Haxton, and Shunsaku Horiuchi for the stimulating conversations.  
This work was supported by U.S. National Science Foundation under grant number PHY-1630782 Focused Research Hub in Theoretical Physics: Network for Neutrinos, Nuclear Astrophysics, and Symmetries (N3AS) (X.W.), and it was also supported by the U.S. Department of Energy under Nuclear Theory Contract No. DE-FG02-95-ER40934 (R.S.), DE-AC52-07NA27344 for the topical collaboration Fission In \textit{R}-process Elements (FIRE; N.V. and R.S.),  and the SciDAC collaborations TEAMS DE-SC0018232 (T.S., R.S.).  
This work benefited from conversations stimulated by the National Science Foundation under grant No. PHY-1430152 (JINA Center for the Evolution of the Elements). 
M.M. was supported by the US Department of Energy through the Los Alamos National Laboratory. Los Alamos National Laboratory is operated by Triad National Security, LLC, for the National Nuclear Security Administration of U.S.\ Department of Energy (Contract No.\ 89233218CNA000001). T.S. was supported in part by the Los Alamos National Laboratory Center for Space and Earth Science, which is funded by its Laboratory Directed Research and Development program under project number No. 20180475DR. M.M. was also supported by the Laboratory Directed Research and Development program of Los Alamos National Laboratory under project number 20190021DR.

{
\software{TALYS \citep[version 1.9,][]{TALYS,TALYS2}, Matplotlib \citep[][http://dx.doi.org/10.1109/MCSE.2007.55]{matplotlib}, Numpy \citep[][https://doi.org/10.1109/MCSE.2011.37]{numpy1, numpy2} }
}

\appendix

\section{A thick-target Solution to the \textit{r}-process Ejecta Propagation}
\label{sec: equations}

For a thick-target limit, the solution to Eq~\ref{eq:prop} with the source function Eq~\ref{eq:source}
gives 
\beqar
\label{eq:ejecta}
N_{\rm i} \left(E,t\right) &=& \frac{1}{b_i\left(n_{\rm gas},E\right)} \int_{E}^{\infty} dE' q_{i,\rm E}\left(E',t'\right)
\nonumber\\
& = & \frac{1}{b_i\left(n_{\rm gas},E\right)} \int_{t}^{\infty} dt' b((n_{\rm gas},E'(t')) N'_{i,0}(E'(t'))\delta(t'-t_0),
\eeqar
where 
\beqar
\label{eq:t}
& &dt=\frac{dE}{b((n_{\rm gas},E)} \Rightarrow dE=b(n_{\rm gas},E(t))dt 
\\
\label{eq:t1}
&\Rightarrow& t' = t-\int_{E}^{E'} \frac{dE''}{b(n_{\rm gas},E'')}.
\eeqar
To do the integral, we need to solve Eq~\ref{eq:t1} for $E$ when $t'=t_0$, i.e., finding $E=E_x (t)$ so that $\Delta t=t-t_0=\int_{E_x}^{E_0} {dE''}/{b(n_{\rm gas},E'')}$. Thus the \textit{r}-process propagated particles scan down in energy as $\Delta t=t-t_0$ goes from 0 onwards, until they reach 
{\em $E_{\rm threshold}=E_x(t_f)\sim 5$ MeV at time $t_f$ when the particles are no longer energetic enough to have spallation reactions}. Therefore we can get $N'_{i,0}(E,t)=N_{i,0}\delta(E-E_x(t))$.

Then we can obtain the spectrum for the propagated \textit{r}-process nuclei $i$:
\beq
\label{eq:spectrum}
N_{i} \left(E,t\right) = \frac{b_i((n_{\rm gas},E_x(t)) N'_{i,0}(E_x(t))} {b_i(n_{\rm gas}, E)} ,
\eeq
and the number fraction of the propagated nuclei $i$ at time $t_f$ to the initial projectile $i$ at $t_0$ is:
\beqar
\label{eq:f_prop}
f_{i, \rm prop}&=& \frac{N_{i, \rm prop}}{N_{i,\rm total}}=\frac{\int N_i(E,t_f) dE}{N_{i,0}} 
\nonumber\\
&=& b_i((n_{\rm gas},E_x(t_f)) \frac{1}{b_i(n_{\rm gas}, E_x(t_f))}
\nonumber\\
&=& 1.
\eeqar
As expected, the particle number is conserved during the propagations.

From the propagated spectrum, we can get the product nuclei ${\ell}$ (mass number $A_{\ell}$, charge $Z_{\ell}$) spectrum (in the lab frame) through the nuclear reactions between projectile nuclei $i$ and nuclei $j$ (mass number $A_j$, charge $Z_j$, number-density ratio $y_j=n_j/n_{\rm gas}$ where $n_j$ is the number density of nucleus $j$ and $n_{\rm gas}$ to be the number density of ISM protons) in the ISM: 
\beqar
\label{eq:prod_spectrum_d}
q_{E,ij}^{\ell}(t) &=& n_{j} \sigma_{ij}^{\ell}(E) v(E) N_{i}(E,t). 
\\
\label{eq:prod_number_d}
\frac{dN_{ij}^{\ell}}{dt} (t)&=& \int q_{E,ij}^{\ell}(t) dE = n_{j}\int_{E_x(t)}^{E_{0}} dE v(E) \sigma_{ij}^{\ell}(E) N_{i}(E,t).
\eeqar

Thus, the number fraction of the total spallation-produced nuclei ${\ell}$ at time $t_f$ to the initial projectile $i$ at time $t_0$ is:
\beqar
\label{eq:fi}
f_{i}^{\ell}=\sum_{j} f_{i,j}^{\ell}&=& \sum_{j} \frac{N_{ij}^{\ell}(t_f)}{N_{i,0}}=\sum_{j} \frac{\int_{t_0}^{t_f} dt dN_{ij}^{\ell}/dt}{N_{i,0}}
\nonumber\\
&=&\sum_{j} \frac{n_{j}}{N_{i,0}} \int_{t_0}^{t_f} dt \int_{E_x(t)}^{E_{0}} dE \frac{b_i((n_{\rm gas},E_x(t)) N'_{i,0}(E_x(t)) v(E) \sigma_{ij}^{\ell}(E) } {b_i(n_{\rm gas}, E)}
\nonumber\\
&=&\sum_{j} y_{j}  \int_{t_0}^{t_f} dt  v(E_x(t)) \sigma_{ij}^{\ell}(E_x(t)) n_{\rm gas}
\nonumber\\
&=&\sum_{j} y_{j} \int_{E_x(t_f)}^{E_0} \frac{\sigma_{ij}^{\ell}(E') v(E') dE'} {b_i(n_{\rm gas},E')/n_{\rm gas}} 
\eeqar
where $E_{0}$ is the kinetic energy per nucleon of the initial projectile nuclei $i$, which is the maximum kinetic energy of the nuclei. 
And $\sigma_{ij}^{\ell}(E)$ is the cross section for the production of nuclei ${\ell}$ by the reaction between ejecta nuclei $i$ and ISM nuclei $j$.

\section{Energy-loss Rate of the \textit{r}-process Nuclei}
\label{sec: b}

With the spallation cross sections shown in Section~\ref{subsec: spall}, we can get the inelastic/spallation energy-loss rate self-consistently (assuming the loss is approximated to be continuous): 
\beq
\label{eq:inelasticb}
b_{\rm spall}^i= v(E) E  \sum_{j} n_{j} \sigma_{ij}^{\rm total}= n_{\rm gas } v(E) E  \sum_{j} y_{j} \sigma_{ij}^{\rm total}.
\eeq
Here, the cross section $\sigma_{ij}^{\rm total}$ is for all of the spallation reactions between projectile nucleus $i$ and target ISM nucleus $j$.
For example, $b_{\rm spall}^{^{196}{\rm Pt}}\sim 2.713\times 10^{-9}\ {\rm MeV/s}$, so the spallation time scale is $\tau_{\rm spall}^{^{196}{\rm Pt}}\sim 196E/b_{\rm spall}^{^{196}{\rm Pt}}\sim 1.76 \rm{Myr}$, for nuclei $^{196}{\rm Pt}$ traveling with a speed of $v=0.3c$ through the ISM with $n_{\rm gas}=1 {\rm cm^{-3}}$, and the spallation mean free path of this nuclei is $\sim 100$kpc.

We also include the energy-loss rate of \textit{r}-process nuclei due to interactions with ISM \citep[or stopping power; the ionization energy loss is the dominant loss channel;][]{Schlickeiser2013}:
\beqar
\label{eq:bi}
b_{\rm ionic} &\sim& 1.82\times10^{-13} Z_{\rm eff}^2 (\frac{n_{\rm gas}}{cm^{-3}})
\nonumber\\
& &  \times(1+0.0185\ln(\beta) H(\beta-\beta_0))\times \frac{2\beta^2}{\beta_0^3+2\beta^3} \ \rm {MeV/A s^{-1}}
\eeqar  
where 
$\beta=v/c$, $H$ is the Heaviside step function, $\beta_0=0.01$ is the orbital velocity of electrons in hydrogen atoms. $Z_{\rm eff}=Z(1-1.034\exp(-137\beta Z^{-0.688}))$ is the effective charge \citep{Brown1972}, which is less than the nucleus's charge $Z$ at low energies.
Then, the ionization loss time scale $\tau_{\rm ionic}^{^{196}{\rm Pt}}\sim E/b_{\rm ionic} \sim 0.1$ Myr for the same condition above, and the mean free path of this nuclei is $\sim 10$kpc. The ionization loss dominate over spallation during the propagation of the \textit{r}-process ejecta. 

Therefore, the total energy-loss rate per nucleon for nucleus $i$ is $b_i(n_{\rm gas},E)=b_{\rm spall}^i/A_i+b_{\rm ionic}^i$. Notice that {\em the energy-loss rate scales with gas density}:  
$b \propto n_{\rm gas}$ (see Eqs~\ref{eq:inelasticb} and \ref{eq:bi}).


\begin{thebibliography}{}

\bibitem[Abbott et al.(2017a)]{Abbott} Abbott, B.~P., Abbott, R., Abbott, T.~D., et al.\ 2017, Physical Review Letters, 119, 161101 

\bibitem[Abbott et al.(2017b)]{NSM} Abbott, B.~P., Abbott, R., Abbott, T.~D., et al.\ 2017, \apjl, 848, L12

\bibitem[Arnould et al.(2007)]{solar2007} Arnould, M., Goriely, S., \& Takahashi, K.\ 2007, \physrep, 450, 97 

\bibitem[Audi et al.(2007)]{Nubase} Audi, G., Kondev, F.~G., Wang, M., Huang, W.~J. \& Naimi, S.\ 2017, Chinese Physics C, 41, 030001

\bibitem[Bauswein et al.(2013)]{Bauswein2013} Bauswein, A., Goriely, S., \& Janka, H.-T.\ 2013, \apj, 773, 78 

\bibitem[Bao et al.(2000)]{NS1} Bao, Z.~Y., Beer, H., K{\"a}ppeler, F., et al.\ 2000, Atomic Data and Nuclear Data Tables, 76, 70 

\bibitem[Berger et al.(2013)]{GRB1} Berger, E., Fong, W., \& and Chornock, R., \ 2013  \apjl, 774, L23.

\bibitem[Binns et al.(2019)]{Binns2019} Binns, W., Israel, M.~H., Rauch, B.~F., et al.\ 2019, \baas, 51, 313

\bibitem[Bovard et al.(2017)]{Bovard} Bovard, L., Martin, D., Guercilena, F., et al.\ 2017, \prd, 96, 124005 

\bibitem[Brodzinski et al.(1971)]{spal10} Brodzinski, R.~L., Rancitelli, L.~A., Cooper, J.~A., \& Wogman, N.~A.\ 1971, \prc, 4, 1250 

\bibitem[Brown \& Moak(1972)]{Brown1972} Brown, M.~D., \& Moak, C.~D.\ 1972, \prb, 6, 90 

\bibitem[Burbidge et al.(1957)]{Burbidge1957} Burbidge, E.~M., Burbidge, G.~R., Fowler, W.~A., \& Hoyle, F.\ 1957, Reviews of Modern Physics, 29, 547

\bibitem[Caballero et al.(2012)]{Caballero2012} Caballero, O.~L., McLaughlin, G.~C., \& Surman, R.\ 2012, \apj, 745, 170

\bibitem[Cameron(1957)]{Cameron1957} Cameron, A.~G.~W.\ 1957, \aj, 62, 9 

\bibitem[Chen \& Beloborodov(2007)]{Chen2007} Chen, W.-X., \& Beloborodov, A.~M.\ 2007, \apj, 657, 383 

\bibitem[Chevalier(1977)]{SNR} Chevalier, R.~A.\ 1977, \araa, 15, 175

\bibitem[Cline \& Nieschmidt(1971)]{spal9} Cline, J.~E., \& Nieschmidt, E.~B.\ 1971, Nuclear Physics A, 169, 437 

\bibitem[Cowan et al.(1991)]{Cowan1991} Cowan, J.~J., Thielemann, F.-K., \& Truran, J.~W.\ 1991, \physrep, 208, 267 

\bibitem[Cowan et al.(2002)]{Cowan2002} Cowan, J.~J., Sneden, C., Burles, S., et al.\ 2002, \apj, 572, 861 

\bibitem[Cowan et al.(2019)]{Cowan2019} Cowan, J.~J., Sneden, C., Lawler, J.~E., et al.\ 2019, arXiv:1901.01410 

\bibitem[Cowperthwaite et al.(2017)]{Cowperthwaite2017} Cowperthwaite, P.~S., Berger, E., Villar, V.~A., et al.\ 2017, \apjl, 848, L17 

\bibitem[Dessart et al.(2009)]{Dessart2009} Dessart, L., Ott, C.~D., Burrows, A., Rosswog, S., \& Livne, E.\ 2009, \apj, 690, 1681 

\bibitem[Duncan et al.(1992)]{Duncan1992} Duncan, D.~K., Lambert, D.~L., \& Lemke, M.\ 1992, \apj, 401, 584

\bibitem[Endrizzi et al.(2016)]{Endrizzi2016} Endrizzi, A., Ciolfi, R., Giacomazzo, B., Kastaun, W., \& Kawamura, T.\ 2016, Classical and Quantum Gravity, 33, 164001 

\bibitem[Fields et al.(1994)]{Fields1994} Fields, B.~D., Olive, K.~A., \& Schramm, D.~N.\ 1994, \apj, 435, 185.


\bibitem[Fields et al.(2000)]{Fields2000} Fields, B.~D., Olive, K.~A., Vangioni-Flam, E., \& Cass{\'e}, M.\ 2000, \apj, 540, 930

\bibitem[Fields et al.(2002)]{Fields2002} Fields, B.~D., Daigne, F., Cass{\'e}, M., \& Vangioni-Flam, E.\ 2002, \apj, 581, 389

\bibitem[Fink et al.(1987)]{spal5} Fink, D., Paul, M., Hollos, G., et al.\ 1987, Nuclear Instruments and Methods in Physics Research B, 29, 275 

\bibitem[Frebel et al.(2007)]{Frebel2007} Frebel, A., Christlieb, N., Norris, J.~E., et al.\ 2007, \apjl, 660, L117 

\bibitem[Foucart et al.(2015)]{Foucart2015} Foucart, F., O'Connor, E., Roberts, L., et al.\ 2015, \prd, 91, 124021

\bibitem[Garrett \& Turkevich(1973)]{spal8} Garrett, C.~K., \& Turkevich, A.~L.\ 1973, \prc, 8, 594 

\bibitem[George et al.(2009)]{George2009} George, J.~S., Lave, K.~A., Wiedenbeck, M.~E., et al.\ 2009, \apj, 698, 1666 

\bibitem[Goriely et al.(2011)]{Goriely2011} Goriely, S., Bauswein, A., \& Janka, H.-T.\ 2011, \apj, 738, L32

\bibitem[Goriely et al.(2013)]{Goriely2013} Goriely, S., Sida, J.-L., \& Lema\^{\i}tre, J.-F. \ 2013, \prl, 111, 242502

\bibitem[Goriely et al.(2015)]{Goriely2015} Goriely, S., Bauswein, A., Just, 0., Pllumbi, E., \& Janka, H.-Th.,\ 2015, MNRAS, 452, 3894

\bibitem[Hunter (2007)]{matplotlib} Hunter, J.~D.,\ 2007, IEEE Computing in Science \& Engineering, 9 , 3

\bibitem[Higdon et al.(1998)]{Higdon1998} Higdon, J.~C., Lingenfelter, R.~E., \& Ramaty, R.\ 1998, \apjl, 509, L33 

\bibitem[Hill et al.(2002)]{Hill2002} Hill, V., Plez, B., Cayrel, R., et al.\ 2002, \aap, 387, 560 

\bibitem[Hohenberg \& Rowe(1970)]{spal11} Hohenberg, C.~M., \& Rowe, M.~W.\ 1970, \jgr, 75, 4205 

\bibitem[Holmbeck et al.(2019)]{Erika} Holmbeck, E.~M., Sprouse, T.~M., Mumpower, M.~R., et al.\ 2019, \apj, 870, 23 

\bibitem[Holmbeck et al.(2018)]{Erika2018} Holmbeck, E.~M., Beers, T.~C., Roederer, I.~U., et al.\ 2018, \apjl, 859, L24 

\bibitem[Hotokezaka et al.(2013)]{Hotokezaka2013} Hotokezaka, K., Kiuchi, K., Kyutoku, K., et al.\ 2013, \prd, 87, 024001


\bibitem[Ivans et al.(2006)]{Ivans2006} Ivans, I.~I., Simmerer, J., Sneden, C., et al.\ 2006, \apj, 645, 613 

\bibitem[Jin et al.(2015)]{GRB2} Jin, Z.-P.,Li, X., Cano, Z., et al.\ 2015,  \apjl, 811, L22

\bibitem[Jin et al.(2016)]{GRB3} Jin, Z.-P., Hotokezaka, K., Li, X., et al.\ 2016, Nature Communications, 7, 12898 

\bibitem[Jin et al.(2020)]{Jin2019} Jin, Z.-P., Covino, S., Liao, N.-H., et al.\ 2020, Nature Astronomy, 4, 77

\bibitem[Just et al.(2015)]{Just2015} Just, O., Bauswein, A., Ardevol Pulpillo, R., et al.\ 2015, \mnras, 448, 541.

\bibitem[Kajino \& Mathews(2017)]{Kajino2017} Kajino, T., \& Mathews, G.~J.\ 2017, Reports on Progress in Physics, 80, 084901 

\bibitem[Kajino et al.(2019)]{Kajino2019} Kajino, T., Aoki, W., Balantekin, A.~B., et al.\ 2019, Progress in Particle and Nuclear Physics, 107, 109

\bibitem[Kasen et al.(2017)]{Kasen2017} Kasen, D., Metzger, B., Barnes, J., et al.\ 2017, \nat, 551, 80.

\bibitem[Kawano et al.(2016)]{HF} Kawano, T., Capote, R., Hilaire, S., et al.\ 2016, \prc, 94, 014612

\bibitem[Kodama \& Takahashi(1975)]{KT} Kodama, T., \& Takahashi, K.\ 1975, Nuclear Physics A, 239, 489 

\bibitem[Kolsky \& Karol(1993)]{spal3} Kolsky, K.~L., \& Karol, P.~J.\ 1993, \prc, 48, 236

\bibitem[Komiya \& Shigeyama(2017)]{Komiya2017} Komiya, Y., \& Shigeyama, T.\ 2017, \apj, 846, 143.

\bibitem[Koning \& Rochman(2012)]{TALYS} Koning, A.~J., \& Rochman, D.\ 2012, Nuclear Data Sheets, 113, 2841 

\bibitem[Koning \& Rochman(2019)]{TALYS2} Koning, A.~J., \& Rochman, D.\ 2019, Nuclear Data Sheets, 155, 1 

\bibitem[Korobkin et al.(2012)]{Korobkin2012} Korobkin, O., Rosswog, S., Arcones, A., et al.\ 2012, \mnras, 426, 1940


\bibitem[Kusakabe \& Mathews(2018)]{Kusakabe2018} Kusakabe, M., \& Mathews, G.~J.\ 2018, \apj, 854, 183 

\bibitem[Kyutoku et al.(2018)]{Kyutoku2018} Kyutoku, K., Kiuchi, K., Sekiguchi, Y., et al.\ 2018, \prd, 97, 023009


\bibitem[Lehner et al.(2016)]{Lehner2016} Lehner, L., Liebling, S.~L., Palenzuela, C., et al.\ 2016, Classical and Quantum Gravity, 33, 184002 

\bibitem[Lemoine et al.(1998)]{Lemoine1998} Lemoine, M., Vangioni-Flam, E., \& Cass{\'e}, M.\ 1998, \apj, 499, 735  

\bibitem[Letaw et al.(1983)]{Letaw1983} Letaw, J.~R., Silberberg, R., \& Tsao, C.~H.\ 1983, \apjs, 51, 271 

\bibitem[Li, \& Paczy{\'n}ski(1998)]{Li1998} Li, L.-X., \& Paczy{\'n}ski, B.\ 1998, \apj, 507, L59

\bibitem[Longair(1981)]{CR2} Longair M. S., 1981, High energy Astrophysics, 1st edn., Cambridge Univ. Press, Cambridge and New York, p. 420

\bibitem[Lodders(2003)]{Lodders2003} Lodders, K.\ 2003, \apj, 591, 1220

\bibitem[Malkus et al.(2016)]{Malkus2016} Malkus, A., McLaughlin, G.~C., \& Surman, R.\ 2016, \prd, 93, 045021

\bibitem[Mannheim, \& Schlickeiser(1994)]{Mannheim1994} Mannheim, K., \& Schlickeiser, R.\ 1994, \aap, 286, 983

\bibitem[Marketin et al.(2016)]{MT} Marketin, T., Huther, L., \& Mart{\'\i}nez-Pinedo, G.\ 2016, \prc, 93, 025805

\bibitem[Martin et al.(2015)]{Martin2015} Martin, D., Perego, A., Arcones, A., et al.\ 2015, \apj, 813, 2 

\bibitem[McLaughlin \& Surman(2005)]{McLaughlin2005} McLaughlin, G.~C., \& Surman, R.\ 2005, \nphysa, 758, 189

\bibitem[Meneguzzi et al.(1971)]{CR1} Meneguzzi, M., Audouze, J., \& Reeves, H.\ 1971, \aap, 15, 337 

\bibitem[Michel et al.(1997)]{spal2} Michel, R., Bodemann, R., Busemann, H., et al.\ 1997, Nuclear Instruments and Methods in Physics Research B, 129, 153

\bibitem[M{\"o}ller et al.(2003)]{Moller2003} M{\"o}ller, P., Pfeiffer, B., \& Kratz, K.-L.\ 2003, \prc, 67, 055802

\bibitem[M{\"o}ller et al.(2016)]{FRDM} M{\"o}ller, P., Sierk, A.~J., Ichikawa, T., et al.\ 2016, Atomic Data and Nuclear Data Tables, 109, 1

\bibitem[M{\"o}ller et al.(2019)]{Moller2019} M{\"o}ller, P., Mumpower, M., Kawano, T., \& Myers, W.~D.\ 2019, Atomic Data and Nuclear Data Tables, 125, 192

\bibitem[Mumpower et al.(2019)]{Mumpower2019} Mumpower, M.~R., Jaffke, P., Verriere, M., et al.\ 2019, arXiv e-prints, arXiv:1911.06344

\bibitem[Mumpower et al.(2016)]{LANL} Mumpower, M.~R., Kawano, T., \& M{\"o}ller, P.\ 2016, \prc, 94, 064317

\bibitem[Mumpower et al.(2018)]{Matt} Mumpower, M.~R., Kawano, T., Sprouse, T.~M., et al.\ 2018, \apj, 869, 14

\bibitem[Mumpower et al.(2017)]{Matt2017} Mumpower, M.~R., Kawano, T., Ullmann, J.~L., Krti{\v c}ka, M., \& Sprouse, T.~M.\ 2017, \prc, 96, 024612 

\bibitem[Mumpower et al.(2016)]{Matt2016} Mumpower, M.~R., Surman, R., McLaughlin, G.~C., et al.\ 2016, Progress in Particle and Nuclear Physics, 86, 86.

\bibitem[Nakamura \& Shigeyama(2004)]{Nakamura2004} Nakamura, K., \& Shigeyama, T.\ 2004, \apj, 610, 888 

\bibitem[Oliphant(2006)]{numpy1} Oliphant, T.~E.,\ 2006,  A Guide to NumPy, USA: Trelgol Publishing  

\bibitem[Paradela et al.(2017)]{spal00} Paradela, C., Tassan-Got, L., Benlliure, J., et al.\ 2017, \prc, 95, 044606

\bibitem[Perego et al.(2014)]{Perego2014} Perego, A., Rosswog, S., Cabez{\'o}n, R.~M., et al.\ 2014, \mnras, 443, 3134 

\bibitem[Perron(1976)]{spal7} Perron, C.\ 1976, \prc, 14, 1108 

\bibitem[Placco et al.(2017)]{Placco2017} Placco, V.~M., Holmbeck, E.~M., Frebel, A., et al.\ 2017, \apj, 844, 18 

\bibitem[Ramaty et al.(2000)]{Ramaty2000} Ramaty, R., Scully, S.~T., Lingenfelter, R.~E., \& Kozlovsky, B.\ 2000, \apj, 534, 747 

\bibitem[Rauscher, \& Thielemann(2001)]{NS2} Rauscher, T., \& Thielemann, F.-K.\ 2001, Atomic Data and Nuclear Data Tables, 79, 47

\bibitem[Rauscher(2010)]{NS3} Rauscher, T.\ 2010, \prc, 81, 045807

\bibitem[Radice et al.(2018)]{Radice18} Radice, D., Perego, A., Hotokezaka, K., et al., 2018, \apj, 869, 130

\bibitem[\protect\citeauthoryear{Reeves, Fowler \& Hoyle}{1970}]{Reeves1970} Reeves H., Fowler W.~A., Hoyle F., 1970, Natur, 226, 727

\bibitem[Regnier(1979)]{spal6} Regnier, S.\ 1979, \prc, 20, 1517 

\bibitem[Rejmund et al.(2001)]{spal0} Rejmund, F., Mustapha, B., Armbruster, P., et al.\ 2001, \nphysa, 683, 540

\bibitem[Roederer et al.(2014)]{Roederer2014} Roederer, I.~U., Preston, G.~W., Thompson, I.~B., et al.\ 2014, \apj, 784, 158

\bibitem[Rosswog et al.(2013)]{Rosswog2013} Rosswog, S., Piran, T., \& Nakar, E.\ 2013, \mnras, 430, 2585

\bibitem[Rosswog et al.(2017)]{Rosswog2017} Rosswog, S., Feindt, U., Korobkin, O., et al.\ 2017, Classical and Quantum Gravity, 34, 104001

\bibitem[Rosswog et al.(2018)]{Rosswog2018} Rosswog, S., Sollerman, J., Feindt, U., et al.\ 2018, \aap, 615, A132 

\bibitem[Sekiguchi et al.(2016)]{Sekiguchi2016} Sekiguchi, Y., Kiuchi, K., Kyutoku, K., Shibata, M., \& Taniguchi, K.\ 2016, \prd, 93, 124046

\bibitem[Schlickeiser(2002)]{Schlickeiser2013} Schlickeiser, R.\ 2013, Cosmic Ray Astrophysics (Berlin: Springer)

\bibitem[Schmidt et al.(2016)]{GEF} Schmidt, K.-H., Jurado, B., Amouroux, C., \& Schmitt, C.\ 2016, Nuclear Data Sheets, 131, 107 

\bibitem[Schmitt et al.(2014)]{Schmitt2014} Schmitt, C., Schmidt, K.-H., \& Keli\ifmmode \acute{c}\else \'{c}\fi{}-Heil, A.\ 2014, \prc, 90, 064605

\bibitem[Schmitt et al.(2016)]{Schmitt2016} Schmitt, C., Schmidt, K.-H., \& Keli\ifmmode \acute{c}\else \'{c}\fi{}-Heil, A.\ 2016, \prc, 94, 039901

\bibitem[Shibagaki et al.(2016)]{Shibagaki2016} Shibagaki, S., Kajino, T., Mathews, G.~J., et al.\ 2016, \apj, 816, 79

\bibitem[Siegel \& Metzger(2018)]{Siegel2018} Siegel, D.~M., \& Metzger, B.~D.\ 2018, \apj, 858, 52 

\bibitem[Sneden et al.(2003)]{Sneden2003} Sneden, C., Cowan, J.~J., Lawler, J.~E., et al.\ 2003, \apj, 591, 936

\bibitem[Sneden et al.(2008)]{Sneden2008} Sneden, C., Cowan, J.~J., \& Gallino, R.\ 2008, \araa, 46, 241 


\bibitem[Surman et al.(2006)]{Surman2006} Surman, R., McLaughlin, G.~C., \& Hix, W.~R.\ 2006, \apj, 643, 1057.

\bibitem[Surman et al.(2008)]{Surman2008} Surman, R., McLaughlin, G.~C., Ruffert, M., Janka, H.-T., \& Hix, W.~R.\ 2008, \apjl, 679, L117 

\bibitem[Suzuki \& Yoshii(2001)]{Suzuki2001} Suzuki, T.~K., \& Yoshii, Y.\ 2001, \apj, 549, 303 


\bibitem[Tanaka \& Hotokezaka(2013)]{Tanaka2013} Tanaka, M., \& Hotokezaka, K.\ 2013, \apj, 775, 113 

\bibitem[Thielemann et al.(2017)]{Thielemann2017} Thielemann, F.-K., Eichler, M., Panov, I.~V., et al.\ 2017, Annual Review of Nuclear and Particle Science, 67, 253

\bibitem[Tobin \& Karol(1989)]{spal4} Tobin, M.~J., \& Karol, P.~J.\ 1989, \prc, 39, 2330 


\bibitem[van der Walt et al.(2011)]{numpy2} van der Walt, S., Colbert, S.~C., \& Varoquaux, G.\ 2011, Computing in Science and Engineering, 13, 22

\bibitem[Vassh et al.(2019a)]{Nicole} Vassh, N., Vogt, R., Surman, R., et al.\ 2019, Journal of Physics G Nuclear Physics, 46, 065202

\bibitem[Vassh et al.(2019b)]{Nicole1} Vassh, N., Mumpower, M.~R., McLaughlin, G.~C., et al.\ 2019, arXiv e-prints, arXiv:1911.07766


\bibitem[\protect\citeauthoryear{Walker, Viola \& Mathews}{1985}]{Walker1985} Walker T.~P., Viola V.~E., Mathews G.~J., 1985, ApJ, 299, 745

\bibitem[Wanajo et al.(2014)]{Wanajo2014} Wanajo, S., Sekiguchi, Y., Nishimura, N., et al.\ 2014, \apj, 789, L39

\bibitem[Wang et al.(2017)]{AME} Wang, M., Audi, G., Kondev, F.~G., et al.\ 2017, Chinese Physics C, 41, 030003

\bibitem[Wang, \& Fields(2018)]{Wang2018} Wang, X., \& Fields, B.~D.\ 2018, \mnras, 474, 4073


\bibitem[Watson et al.(2019)]{Watson2019} Watson, D., Hansen, C.~J., Selsing, J., et al.\ 2019, \nat, 574, 497

\bibitem[Wollaeger et al.(2018)]{Wollaeger2018} Wollaeger, R.~T., Korobkin, O., Fontes, C.~J., et al.\ 2018, \mnras, 478, 3298

\bibitem[Yashima et al.(2002)]{spal1} Yashima, H., Uwamino, Y., Sugita, H., et al.\ 2002, \prc, 66, 044607


\end{thebibliography}
\end{document}